\RequirePackage{graphicx}
\documentclass[aps,pra,floatfix,preprint, longbibliography]{revtex4-1}
\usepackage{amssymb}
\usepackage{enumitem}
\usepackage{mathtools}
\usepackage{xcolor}

\usepackage{float}

\usepackage{graphicx}
\usepackage{natbib}
\usepackage{dcolumn}
\usepackage{bm}
\usepackage{mathtools}
\usepackage{braket}
\usepackage[utf8x]{inputenc}
\usepackage{amsmath}
\usepackage{multirow}
\usepackage{amsfonts}
\usepackage{changepage}
\usepackage{amssymb}
\usepackage{rotating} 

\usepackage{amsthm}

\theoremstyle{plain}
\theoremstyle{definition}

\begin{document}
\title{Lowering Tomography Costs in Quantum Simulation with a Symmetry Projected Operator Basis }



\author{Scott E. Smart and David A. Mazziotti}
\email[]{damazz@uchicago.edu}
\affiliation{Department of Chemistry and The James Franck Institute, The University of Chicago, Chicago, IL 60637}%

\date{Submitted 7 August 2020, Published January 28, 2021. Phys. Rev. A 103, 012420.}

\begin{abstract}
Measurement in quantum simulations provides a means for extracting meaningful information from a complex quantum state, and for quantum computing, reducing the complexity of measurement will be vital for near-term applications. For most quantum simulations, the targeted state will obey several symmetries inherent to the system Hamiltonian. We obtain an alternative symmetry projected basis of measurement that reduces the number of measurements needed \textcolor{black}{by a constant factor}. Our scheme can be implemented at no additional cost on a quantum computer, can be implemented under various measurement or tomography schemes, and is reasonably resilient under noise.
\end{abstract}
\baselineskip24pt

 \maketitle

\section{Introduction}
One of the fundamental challenges in quantum simulation is the storage and propagation of exponentially scaling many-body quantum states. Many classical computational methods treat these states approximately, using perturbative and truncated approaches or local approximations, and result in polynomial algorithms which potentially sacrifice key characteristics of the quantum state~\cite{Szabo1996, Helgaker2000}. Reduced density matrix (RDM) methods focus on reducing the required state information to the $k-$body interaction inherent in the system (such as the 2-RDM for fermionic simulations)~\cite{Coleman2000, Mazziotti2007, Zhao2004, Mazziotti2004a, Shenvi2010, Mazziotti2011, Verstichel2012, Schilling2013, Mazziotti2016, Piris2017a, Rubio-Garcia2019, Mazziotti2006, Smart2020a}.  In these cases, the 2-RDM then is subject to its own set of criteria, such as \textcolor{black}{$N-$}representability, but can deal with many-body phenomena in an easier manner~\cite{Coleman2000, Coleman1963, Mazziotti2012, Boyn2020}.

\textcolor{black}{An easy way to reduce the required amount of information to describe the state is through utilizing symmetries.  Symmetries are conserved quantities, preserved through the preparation and propagation of a state}~\cite{Feynman1963, Bishop1993, Helgaker2000}. For molecular systems, particle number, total and projected spin, invariance under time-reversal, and often molecular point groups are examples of symmetries. These can be used in classical electronic structure calculations to \textcolor{black}{substantially} reduce the number of resources necessary to simulate a system \cite{Helgaker2000,Gidofalvi2005,Nakatsuji1979}. On a quantum computer, the exponentially scaling state can be prepared efficiently for many applications. For systems with a $k-$body interaction, tomography of the $k-$RDM is sufficient to describe the system's physical properties. Storage of the state (which is critical for near-term applications) can be reduced to the tomography of the $k$-RDM based on the $k-$body interaction~\cite{Rubin2018, Bonet-Monroig2019, Smart2019,  Sager2020, Smart2020a}.  \textcolor{black}{In contrast to classical RDMs method, a pure quantum state would automatically satisfy the $N-$representability problem.}

 \textcolor{black}{In quantum computing, symmetries are utilized in a wide range of settings. There is ongoing work to utilize number-preserving gate sequences \cite{Roth2017,Ganzhorn2019},
design algorithms for preparing symmetry preserving ansatz \cite{Wang2009,Whitfield2013,
Barkoutsos2018,Fischer2019,Barron2020,Gard2020,Smart2020}, 
reduce symmetry violations through variational constraints \cite{McClean2016, Kandala2017,Ryabinkin2019},
reduce the simulated Hilbert space through varied applications of symmetries \cite{Moll2016,Bravyi2017,Setia2019}, 
and to utilize symmetries as a form of error mitigation \cite{McArdle2019, Bonet-Monroig2018, Sagastizabal2019}. }

Despite potential reductions in the state complexity using symmetries or other methods, \textcolor{black}{tomography still can require} a large set of measurements. For molecular systems, the Hamiltonian and 2-RDM scale as $O(r^4)$, where $r$ is the number of basis functions, and many heuristic and systematic ways involving graph-theoretic or combinatorial approaches have been introduced to lower this number, which in some cases can render an apparent scaling of $O(r^3)$ \textcolor{black}{or $O(r^2)$ with swap networks}~\cite{Izmaylov2019, Izmaylov2019b, Bonet-Monroig2019,Gokhale2019}. Within quantum simulation as a whole, fermionic tomography is particularly challenging due to the nonlocal characteristics of fermionic operators and prevents a logarithmic reduction \textcolor{black}{in complexity} seen with other qubit systems~\cite{Bonet-Monroig2019}.

In this work we present a method of lowering measurement and tomography costs for quantum states and RDMs by exploiting the quantum state's symmetries. By finding the symmetry projected form of our measurement operators, we can re-express our operators in a minimal basis on the quantum computer. The method leads to a constant \textcolor{black}{scaling} improvement in the number of terms which has to be measured, and can be combined with other measurement techniques to reduce circuit preparation costs for near-term calculations.
 \section{Theory}
A symmetry for a quantum system can be defined mathematically as a non-zero operator $\hat{S}$ which commutes with the system Hamiltonian $\hat{H}$,
\begin{equation}
[\hat{H},\hat{S}] = 0.
\end{equation}
Consider a set of $n$ symmetries $\mathcal{S}=\{S_1,S_2,...,S_n\}$ where each symmetry commutes with all other symmetries (note that the most common set in fermionic simulation of $\hat{N},\hat{S}_z$, and $\hat{S}^2$ obeys this). We can find a basis which is a mutual eigenbasis of each element of $\mathcal{S}$, and then we denote a wavefunction which obeys each of these symmetries:
\begin{equation}
|\psi \rangle = \sum_\alpha c_\alpha |\alpha,s_1,s_2,...,s_n\rangle,
\end{equation}
where each $s_i$ represents the eigenvalues of the $i$-th symmetry. Now, let  $\hat{A}$ be an operator acting on this state in this symmetry basis:
\begin{equation}
\hat{A} = \sum a^{\alpha,t_1,t_2,...t_n}_{\beta,u_1,u_2,...u_n} | \alpha, t_1,t_2, ...,t_n\rangle \langle \beta, u_1,u_2,...,u_n|.
\end{equation}
Then, if we are interested in the expectation of $\hat{A}$, we can evaluate it as:
\begin{equation}\label{asymm}
\langle \hat{A} \rangle = \sum_{i,j} c^*_i c^{}_j a^{i,s_1,s_2,....,s_n}_{j,s_1,s_2,...,s_n}
\end{equation}
and so we have projected $\hat{A}$ into the specific subspace of each symmetry, despite that $\hat{A}$ does not necessarily commute with $\hat{S}$. Note that if each symmetry did not commute, our eigenvectors would not be simultaneous eigenstates, and we could instead apply the operators in terms of increasing restrictions as relevant to the quantum state. 

In quantum simulation, \textcolor{black}{often we measure operators which likely do not violate these symmetries individually but often cannot} be directly measured on the quantum computer.\textcolor{black}{ Instead, we map these operators to a set of measurable operators on the quantum device, which commonly today are projective measurements onto eigenvectors of the Pauli matrices. The basis of operators will not always commute with the state's symmetries, and so will be projected by the wavefunction.}

\textcolor{black}{To} find the projected form, we could explicitly calculate the operator form in Eq.~\eqref{asymm} for small systems, but this quickly becomes unfeasible with increasing system size. By noting that most \textcolor{black}{operators} we are interested \textcolor{black}{ in act non-trivially on a few local sites, we can find a projected form acting on the local space.} \textcolor{black}{The form of this operator can be found relatively easily for particular symmetries, and we derive our approach in Appendix A.}  Instead of focusing on one \textcolor{black}{symmetry state $s$}, we project \textcolor{black}{our operator onto a symmetry \emph{conserving}} \textcolor{black}{subspace}:
\begin{equation}
\langle \tilde{A} \rangle = \sum_s \textcolor{black}{\langle \hat{P}_s \hat{A} \hat{P}_s \rangle} =\sum_s \sum_{i,j}  c_i^* c_j^{} a^{i,s}_{j,s},
\end{equation}
where $\hat{P}_s$ is a projection \textcolor{black}{onto a single symmetry $s$. The projected form can be viewed as a mixed operator resulting from projecting onto different pure symmetry states}. \textcolor{black}{Notably, both} $\tilde{A}$ and the projected $\hat{A}$ will \textcolor{black}{generally contain} significantly less \textcolor{black}{terms than the native qubit operators}. 

With these points in view, our approach is as follows. Given an operator $\hat{M}$, we express it in the Pauli basis using some transformation:
\begin{equation}
\hat{M} = \sum_i a_i \hat{A}_i 
\end{equation}
where $\hat{A}_i$ are typically Pauli strings. Then, we apply our symmetry projection to the individual $\hat{A}_i^c = \sum_s \hat{P}_s \hat{A}_i \hat{P}_s $. We represent both the operator and the Pauli strings in a vector form ($\vec{m}$, and $\vec{A}_i^c $) and then using $\vec{A}_i^c$ as columns, form a matrix of linearly independent vectors, $U$.

Finally, we solve the linear system of equations for a vector $\vec{x}$:
\begin{equation}
U\vec{x} = \vec{m}
\end{equation}
to obtain a new basis of measurement for $\hat{M}$ which is equal to or lower in dimension. In general, this will not be unique, and we can order our selection process or bias it to affect the set of terms. The process here is summarized in Table 1 \textcolor{black}{and an example is included in Appendix B.}
\begin{table}\caption{Example procedure for finding a set of symmetry projected operators. The vector representation of $\hat{A}$ is given as $\vec{A}$.}
\def\arraystretch{0.75}
\begin{tabular}{l}
\hline\hline
Given operators sets for measurement ($M$), symmetries ($\mathcal{S}$) and the computational basis ($A$);\\
\hline
(0) Find a set of projection operators $\hat{P}_s$; \\
(1) For each measurement operator $\hat{m} \in M$: \\
\hspace{1cm} (a) Transform $\hat{m}$ in $A$, $\hat{m}=\sum_i a_i \hat{A}_i $ \\
\hspace{1cm} (b) For each $\hat{A}_i$, find symmetry projected computational operators $\hat{A}_i^c$ \\
\hspace{1cm} (c) Choose linearly independent $\vec{A}_j$ as columns of $U$ and solve for $U\vec{x}_m = \vec{m}$
(2) Apply further processing with new set of operators $\{ \vec{x}_m\} $ \\
\hline \hline
\end{tabular}
\end{table}
\section{Results and Applications}
\subsection{Application to Reduced Density Operators}
\textcolor{black}{This work's inspiration is centered on molecular and fermionic systems, which only need characterization of the particles pairwise interactions.} These are completely captured in the two-electron reduced density matrix, or 2-RDM, which represents a partial tomography of the quantum state. Elements of the 2-RDM are measured according to:
\begin{equation}
{}^2 D^{i,k}_{j,l} = \langle\psi |a^\dagger_i a^\dagger_k a^{}_j a^{}_l|\psi \rangle
\end{equation}
where $i,j,k,$ and $l$ are spin orbital indices. On the quantum computer, the most basic \textcolor{black}{mapping from fermions to qubits} is the Jordan-Wigner \textcolor{black}{transformation}. This transforms the creation and annihilation operators as~\cite{Jordan1928}:\begin{equation}
a^\dagger_j = \frac{1}{2}(X_j -iY_j ) \bigotimes_{k=1}^{j-1} Z_k ,
\end{equation}
\begin{equation}
a^{}_j = \frac{1}{2}(X_j +iY_j ) \bigotimes_{k=1}^{j-1} Z_k,
\end{equation}
\textcolor{black}{where $X_j$, $Y_j$, and $Z_j$ indicate a Pauli operator acting on qubit $j$. The local aspect of the operation on a qubit is defined by the $X$ and $Y$ gates, whereas the $Z$ portion generates parity-conserving gates}. The parity mapping~\cite{Tranter2015,Tranter2018} exchanges the storage of orbital occupations and parity, and the Bravyi-Kitaev mapping stores both in a tree-like diagram~\cite{Bravyi2000}. Both of these schemes form linear combinations of operators which act differently on local sites and identically on nonlocal sites, and  thus can be symmetry projected with our technique.

For higher order RDMs (which can be used for exploring states in a linear- or quadratic- expansive subspace or in other methods~\cite{McClean2017}), similar advantages can be seen. We show the effect of our symmetry projection technique in reducing the \textcolor{black}{number of measurements for 1, 2, and 3-RDMs} in Table II. \textcolor{black}{We specify cases with different number of excitations because particle and hole operators commute with the given molecular symmetries, whereas excitations (or de-excitations) will transform to operators in the computational basis that \emph{individually} may not conserve a given symmetry.} The set of measurements in performing tomography of the 2-RDM \textcolor{black}{contains the set required for} molecular Hamiltonians, and thus can be used in Hamiltonian based measurement. \textcolor{black}{An advantage of focusing on the 2-RDM is that it enables systematic approaches in the tomography and that the 2-RDM itself can be used as a tool in error-mitigation~\cite{Smart2019,McClean2017, Gokhale2019b,Bonet-Monroig2019,  Smart2020}.}

\begin{table}[h]
\caption{Dimension of the number of Pauli measurements required for tomography of the 1- and 2-RDMs in the traditional (naive) and symmetry projected (reduced) approaches for given spin and spatial configurations of the second quantized measurement operators with $\hat{N}$, and $\hat{S_z}$ symmetries \textcolor{black}{under the Jordan-Wigner transformation}. \textcolor{black}{We also give examples of the sets of operators in the naive ($N$) and reduced ($R$) methods corresponding with operators marked with a $(*)$. Note, $R$ is not unique in any of these cases.} The cases including the $\hat{S}^2$ symmetry do not greatly affect the results, but require \textcolor{black}{covering the many} permutations of the spatial orbitals. A bar across spins indicates an excitation or de-excitation between these orbitals, and only the unique spin configuration is shown. \textcolor{black}{$\times$ indices the Cartesian product of two sets. $\vec{Z}$ indicates a tensor product of $Z$ gates which is constant across all operations.}}
\def\arraystretch{0.8}
\begin{centering}
\begin{tabular}{ccc|cc|c}
$k-$RDM & Spin & $q-$Sites & Naive  & Reduced  & Example Operator and Sets of Measurement Operators \\
\hline\hline
1 &  $\alpha\alpha$ &  1 & 2 & 2 &  \multirow{3}{0.6\linewidth}{ \centering $ a^\dagger_{i} a^{}_{j},~~ \{ i,j\} \in \alpha $ \\$ \textrm{N}= \{X_i,Y_i\}\times \{X_j,Y_j\} \vec{Z} ,~ \textrm{R}  =  \{ X_i X_j\vec{Z}, X_i Y_j \vec{Z} \}$} \\
 & $\bar{\alpha\alpha}^*$ &  2  & 4 & 2\\
 & $\alpha\beta$ & - & - & 0 \\
 \hline \hline
2 & $\alpha\alpha\alpha\alpha$ &  2 & 4 & 4 & \multirow{7}{0.6\linewidth}{ \centering $a^\dagger_{i} a^{\dagger}_{k} a^{}_{l} a^{}_{j}+ a^\dagger_{j} a^{\dagger}_{l} a^{}_{k} a^{}_{i }$ \\$ \{ i,j\} \in \alpha, \{ k,l \} \in \beta $   \\ $ \textrm{N} =  \{ X_{i } X_{k}, Y_{i}Y_{k}\} \times \{X_{l} X_{j},Y_{l} Y_{j}\}\vec{Z} ~\cup   $ \\ $
 \{ X_{i} Y_{k}, Y_{i}X_{k}\} \times \{X_{l} Y_{j},Y_{l}X_{j}\} \vec{Z}  ,
$ \\ 
$\textrm{R} =  \{  X_{i} X_{k  } X_{l} X_{j}\vec{Z} ,X_{i} Y_{k} Y_{l} X_{\j} \vec{Z} \}$ } \\
  & $\alpha\alpha\bar{\alpha\alpha}$ &  3 & 8 & 4\\
  & $\bar{\alpha\alpha}\bar{\alpha\alpha}$ &  4 & 16 & 6 \\
  & $\alpha\alpha\alpha\beta$ & - &  -  & 0 \\
 & $\alpha\alpha\beta\beta$ &  2 & 4 & 4 \\
  & $\alpha\alpha\bar{\beta\beta} $ & 3 & 8 & 4 \\
 & $\bar{\alpha\alpha}\bar{\beta\beta}^* $ & 4 & 16 & 4  \\
 \hline\hline
3 & $\alpha\alpha\alpha\alpha\alpha\alpha$ & 3 & 8 & 8 & \multirow{12}{0.6\linewidth}{ \centering 
$ a^\dagger_{i} a^{\dagger}_{k} a^{\dagger}_{m} a^{}_{n} a^{}_{l} a^{}_{j}+ a^\dagger_{j} a^{\dagger}_{l} a^{\dagger}_{n} a^{}_{m} a^{}_{k} a^{}_{i}  $\\ $ \{i,j,k,l\} \in \alpha,~\{m,n\} \in \beta  $ \\ 
$ \textrm{N} =  \{ X_{i } X_{k}, Y_{i}Y_{k}\}\vec{Z} \times  ( \{ X_{m } X_{n}, Y_{m}Y_{n}\} \times \{X_{l} X_{j},Y_{l} Y_{j}\} ~\cup $ \\
$ \{ X_{m } Y_{n}, Y_{m}X_{n}\} \times \{X_{l} Y_{j},Y_{l} X_{j}\} ) ~\cup   $ \\  
$\{ X_{i } Y_{k}, Y_{i}X_{k}\} \vec{Z} \times ( \{ X_{m } Y_{k}, Y_{m}X_{n}\} \times \{X_{l} X_{j},Y_{l} Y_{j}\}~\cup$ \\$ \{ X_{m } X_{n}, Y_{m}Y_{n}\} \times \{X_{l} Y_{j},Y_{l} X_{j}\} )$,\\
$\textrm{R} =  
\{ \{ (X_{i} X_{k}Y_{l}X_{j}, Y_{i}X_{k}X_{l }X_{j}, X_{i} Y_{k}X_{l}X_{j} \} \times Y_{m} X_{n} \vec{Z}~\cup $ $  
\{X_iX_kX_lX_j, Y_iY_kX_lX_j, Y_iX_kY_lX_j \} \times X_{m} X_{n} \vec{Z} \}$
} \\
  & $\alpha\alpha\alpha\alpha\bar{\alpha\alpha}$ & 4 & 16 & 8\\
    & $\alpha\alpha\bar{\alpha\alpha}\bar{\alpha\alpha}$ & 5 & 32 & 12\\
  & $\alpha\alpha\alpha\alpha\alpha\alpha$ & 6 & 64 & 20\\
  & $\alpha\alpha\alpha\alpha\alpha\beta$ & - & - & 0 \\
  & $\alpha\alpha\alpha\alpha\beta\beta$ & 3 & 8 & 8\\
  & $\alpha\alpha\bar{\alpha\alpha}\beta\beta$ & 4 & 16 & 8\\
  & $\alpha\alpha\alpha\alpha\bar{\beta\beta}$ & 4 & 16 & 8\\
  & $\alpha\alpha\bar{\alpha\alpha}\bar{\beta\beta}$ & 5 & 32 & 8\\
  & $\bar{\alpha\alpha}\bar{\alpha\alpha}\beta\beta$ & 5 & 32 & 12\\
  & $\bar{\alpha\alpha}\bar{\alpha\alpha}\bar{\beta\beta}^*$ & 6 & 64 & 12\\
  & $\alpha\alpha\alpha\beta\beta\beta$ & - & - & 0\\
\hline\hline
\end{tabular}
\end{centering}
\end{table}

One technique to lower measurement costs involves using the qubit-wise commutation relation, which relates Pauli strings which can be concatenated and thus simultaneously measured through local measurement schemes. Finding the optimal grouping is a NP hard problem, but by characterizing the set of measurements as a graph problem connected \textcolor{black}{by this relation}, we can use coloring algorithms to reduce the number of colors (or cliques) needed~\cite{Gokhale2019,Verteletskyi2019}. To consider the advantage in using our projected technique, we can compare the number of cliques obtained with \textcolor{black}{the default tomography to number obtained with our method}. We explore this for \textcolor{black}{obtaining 2-RDMs} of differing sizes in Figure \textcolor{black}{1}. Additionally, the \textcolor{black}{worst-case} scaling of the number of tomography terms \textcolor{black}{is} $\mathcal{O}(r^4)$, and we look at \textcolor{black}{overall} scaling coefficient ($r^n$) under the \textcolor{black}{grouping technique} with both measurement schemes. \textcolor{black}{If one used the maximally commuting method \cite{Gokhale2019,Izmaylov2019b}}, which for the most part finds larger groups of operators \textcolor{black}{that can be simultaneously measured}, the circuit depth scales polynomially with the number of terms in a group, and so our scheme would lead to reductions in the transformation required.

\begin{figure}
\caption{(Left) Ratio of the number of total required terms in the 2-RDM versus the number of \textcolor{black}{required circuits following a grouping procedure}, and (\textcolor{black}{right}) the scaling coefficient \textcolor{black}{of the number of circuits} with respect to the number of qubits. \textcolor{black}{The} color denotes the fermionic mapping (Jordan-Wigner, Parity, and Bravyi-Kitaev), and the symbol denotes the set of symmetries applied \textcolor{black}{in the projection procedure}. \textcolor{black}{The grouping procedure involves grouping terms according to qubit-wise commutation following Ref. \cite{Verteletskyi2019}.} The black line on the right refers to total number of terms. \textcolor{black}{See Appendix C for more details.}}
\hspace*{-1cm}\includegraphics[scale=0.75]{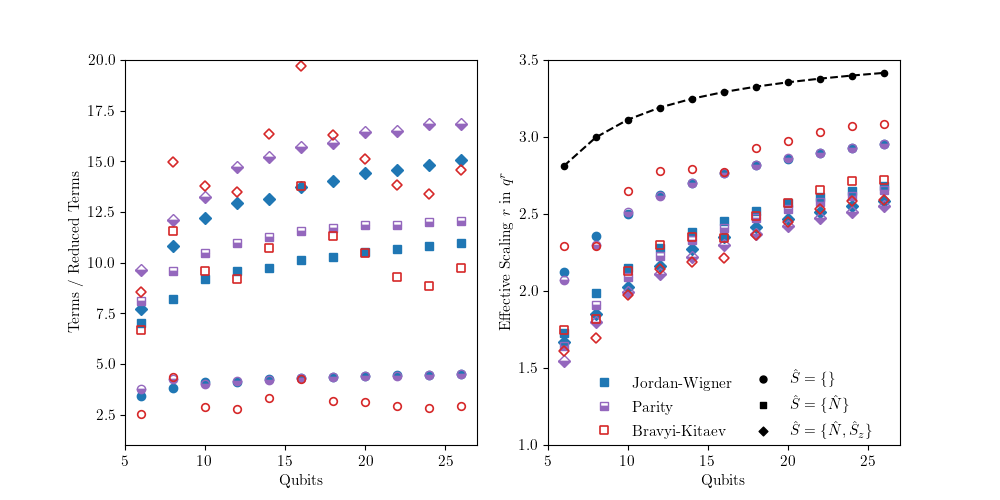}
\end{figure}

\subsection{Effects of Noise on Particle Count}
One implicit assumption in the above work is that the quantum state is of decent quality and that it \textcolor{black}{preserves} the proper symmetries throughout \textcolor{black}{the simulation}. Due to noise, this will almost certainly never be the case, so we are interested in how noise can affect our symmetry projected scheme's quality.

From a theoretical perspective, we can envision two broad cases. In the first, we have a mixed state which is a sum of weighted states:
\begin{equation}
\rho = \sum_s \sum_i  \alpha_{si}|i,s \rangle \langle i,s|.
\end{equation}
In this case the symmetry projection is still exact, as the states are orthogonal to each other, and our method will not be affected by errors. The second case involves a state that is a mixture of different symmetry states, in which case our reduced tomography no longer represents the true tomography. Yet, whether or not standard tomography would offer significant advantages in this case is unclear, as significant errors still corrupt the system in a variety of ways.

\textcolor{black}{To investigate the effects of noise, we simulated a minimal two-fermion in four-spin orbital system on a quantum device and with an accompanying noise model. The results are seen in Table III.} \textcolor{black}{Importantly,} the distance between the \textcolor{black}{RDMs produced by the} tomography methods is comparable to statistical noise, and is always less than then the distance from the true \textcolor{black}{2-RDM, when noise is present}. \textcolor{black}{In the case of the noise-free result, the larger variance between the two tomography methods is likely a result of propagation of sampling error, which is absent with the ideal 2-RDM.}
\begin{table*}[h]
\caption{A comparison of 2-RDMs \textcolor{black}{of a two-electron system} under varying levels of simulated noise \textcolor{black}{(simulated and experimental)} through the Frobenius norm of the difference matrices at randomly sampled points. $^2 D $ refers to the ideal 2-RDM, $^2\tilde{D}$ refers to standard tomography of the 2-RDM under a noise model, and $^2\tilde{D}^c$ refers to the 2-RDM constructed from symmetry projected tomography under a noise model. Values represent averages of the Frobenius norms of difference matrices over 25 random states of H$_2$ in a minimal basis where the ansatz includes 3 parameters. In general, the differences between the noisy tomography methods are consistently much smaller than the difference to the ideal state, and are almost indistinguishable from stochastic effects (seen at the $n=\infty$ limit). \textcolor{black}{More details regarding these results are included in Appendix C.}}
\setlength{\tabcolsep}{8pt}
\def\arraystretch{1.1}
\textcolor{black}{
\begin{tabular}{c|c|c|c}
\hline
$\Delta =$ &   $^2 D_{\rm }-{}^2 \tilde{D} $ & $^2 D_{\rm }-{}^2 \tilde{D}^c $  & \multicolumn{1}{c}{${}^2 \tilde{D}- {}^2 \tilde{D}^c $}  \\
\hline
Noise Strength, $(\frac{1}{2})^n$ & $||\Delta ||_F $ & $||\Delta ||_F$ & $||\Delta ||_F$ \\ \hline
$n=0$ & 0.68(6) & 0.69(5) & 0.05(1)  \\
$n=1$ & 0.39(4) & 0.39(4) & 0.049(9) \\
$n=2$ & 0.22(2) & 0.23(3) & 0.05(1)  \\
$n=3$ & 0.17(1) & 0.17(2) & 0.046(9)  \\
$n=4$ & 0.13(1) & 0.14(1) & 0.05(1) \\
$n=\infty$ & 0.027(5) & 0.036(8) & 0.05(1) \\
\hline Experimental &  0.87(5) & 0.87(4) &  0.05(1) \\
\hline 
\end{tabular}
}
\end{table*}

\section{Conclusion}
Modern quantum computing has advanced drastically in the past decade, with a surge of incremental improvements in experimental and algorithmic improvements. Circuit optimization, qubit reduction, reducing the required parameter space in the classical optimization, or lowering tomography and measurement costs all have attempted to capitalize on the available quantum resources maximally. Work in utilizing system symmetries explores a fascinating aspect of quantum mechanics, and we hope that future work will continue to apply these ideas in lowering costs.

Our approach in utilizing symmetry projected operators provides a simple way to reduce the number of measurements needed, and when combined with other techniques can lead to large reductions in the effective scaling of the system. The routine can be performed in one step before the calculation, and adds no additional cost to the quantum or classical algorithm.
\section*{Acknowledgements} D.A.M. gratefully acknowledges the Department of Energy, Office of Basic Energy Sciences, Grant DE-SC0019215 and the U.S. National Science Foundation Grants No. CHE-2035876 and No. DMR-2037783.

\section*{Data Availability}

Data is available from the corresponding author upon reasonable request.

\appendix
\textcolor{black}{
\section{Derivation of Local Symmetry Operators}}
\textcolor{black}{
In Section II we noted we would like to use a symmetry which exists on the local subspace of the system instead of the entire system for our method. While in general this will not hold for every symmetry, we discuss ways to identify and utilize these symmetries here. 
}

\textcolor{black}{ 
If we consider a state as being in an eigenstate $s$ of a symmetry $\hat{S}$, as seen in Eq. (5) the projection of the state of $s$ onto an operator $A$ is:
\begin{equation}
\langle P_s \hat{A} P_s \rangle = \sum_{i,j} c_{i,s} c^*_{j,s} a^{i,s}_{j,s}. 
\end{equation}
Let the state exist in the Fock space of $M$ orbitals, $\mathcal{F}(M)$. Assume that we have symmetry operators $\hat{T}_1$ and $\hat{T}_2$ which act on subspaces $F_1 = \mathcal{F}(N)$ and $F_2 = \mathcal{F}(M-N)$ respectively, and that we can express an eigenstate of $\hat{S}$ as:
\begin{equation}
|i,s\rangle = \sum_{\alpha,\beta} \sum_{t} R^{i}_{s} {}^{\alpha}_{t} {}^{\beta}_{t'} |\alpha, t  \rangle \otimes |\beta, t' \rangle
\end{equation}
where $t$ is an eigenstate of $\hat{T}_1$, $t'$ is an eigenstate on $\hat{T}_2$ such that the direct product of vectors will be eigenvectors of the symmetry value $s$, and $R$ is a rank-three tensor providing an index between the total space and the two smaller subspaces. We can also imagine a reduced representation, where we omit some of the exact state information, considering only blocks in different symmetries (as the orthogonality of the states is not used here): 
\begin{equation} 
|s \rangle = \sum_{t} R^s_{t,t'}. |t\rangle \otimes |t' \rangle  
\end{equation}.
Now, let $\hat{A}$ be an operator which acts primarily on $F_1$ and only adds a phase onto elements of $F_2$, which in denoted as:
\begin{equation}
\hat{A} = \sum_{a,b,c} A_{a,b,c} | a\rangle \langle  b | \otimes | c \rangle \langle c | 
\end{equation} where $A$ is also a rank-three tensor, indicating the phase for a given state in $F_2$. For non-fermionic operators this would is unity across all states, whereas for fermionic operators there will be some phase changes depending on the states $\gamma$. Using this to evaluate the expectation of $\hat{A}$ we find:
\begin{align}
 {\rm Tr~} \hat{A} \rho =&  {\rm Tr} \sum_{i,j}\sum_{\alpha,\beta,\gamma,\delta,\epsilon,\zeta,\eta}\sum_{a,b,c,d,e}
c_{i,s}^{} c^*_{j,s}
A^{\alpha,\beta,\gamma}_{a,b,c}
R^{}{}^{i,\delta,\epsilon}_{s,d,d’}
R^*  {}^{j,\zeta,\eta}_{s,e,e’}
|\alpha,a \rangle \langle \beta,b| \delta,d \rangle \langle \zeta, e| \\ \notag  &\otimes  
|\gamma,c \rangle \langle \gamma,c | \epsilon,d'\rangle \langle  \eta, e'| \\
=& {\rm Tr}_1 \sum_{i,j} \sum_{\alpha,\beta,\gamma,\delta,\zeta} \sum_{a,b,c}
c_{i,s}^{} c^*_{j,s}
A  {}^{\alpha}_{a} {}^{\beta}_{b} {}^{\gamma}_{c}
R  {}^{i}_{s} {}^{\delta}_{c'} {}^{\gamma}_{c}
R^*{}^{j}_{s} {}^{\zeta}_{c'} {}^{\gamma}_{c} 
|\alpha,a \rangle \langle \beta,b| \delta,c' \rangle \langle \zeta, c'| 
\\ 
=& \sum_{k} {\rm Tr_1}  \sum_{i,j}\sum_{\alpha,\beta,\gamma,\delta,\zeta}
c^{}_{i,s} c^*_{j,s} 
A  {}^{\alpha}_{k} {}^{\beta}_{k} {}^{\gamma}_{k'}
R  {}^{i}_{s} {}^{\delta}_{k} {}^{\gamma}_{k'}
R^*{}^{j}_{s} {}^{\zeta}_{k} {}^{\gamma}_{k'} 
|\alpha,k \rangle \langle \beta,k| \delta,k \rangle \langle \zeta, k| .
\end{align}
In the last two steps we used the fact that following the partial trace over $F_2$ the total spin of the system must still give $s$, yielding a mixed state of pure symmetry states of $\hat{T}_1$. Through linearity we can extract a projection operator and apply it to elements of $\hat{A}$, which leads to a greatly simplified form of the operator. This is equvialent to Eq. (5) in the main text, and this projected form can be found for our set of transformed measurement opeartors. 
}

\textcolor{black}{
The symmetries discussed in the text relating to molecular systems, $\hat{N}$, $\hat{S}_z$, and $\hat{S}^2$, all can be described in this way, and thus can be applied using this method. Symmetries which are elements of the Pauli group (i.e. tensored products of Pauli matrices) also satisfy these conditions. 
}

\textcolor{black}{
\section{Examples of Symmetry Projection}
To illustrate our method, let $|\psi \rangle$ be a two-electron system in four $\alpha-$spin orbitals and $\hat{M}$ be defined as:
\begin{equation}
\hat{M} = c_1 a^\dagger_1 a^{}_{2}+ c_2  a^\dagger_2 a^{}_{1}.
\end{equation}
$\hat{M}$ here is a linear combination of 1-RDM operators. The system as a whole will obey the number symmetry $\hat{N}=\sum_{i=1}^4 a^\dagger_i a^{}_i$, but we can apply a reduced symmetry $\hat{N}_1 = \sum_{i=1}^2 a^\dagger_i a^{}_i$ to our operator. The reduced symmetry operator will have projections operators onto the $N=0$, $N=1$, and $N=2$ subspaces, which can be written as:
\begin{align}
P_0 = |00\rangle \langle 00|,~P_1 = |01 \rangle\langle 01| + |10 \rangle \langle 10|,~P_2 = |11 \rangle \langle 11|.
\end{align}
Using the Jordan-Wigner transformation we can express $\hat{M}$ in terms of Pauli matrices yielding: 
\renewcommand{\arraystretch}{0.6 }
\begin{equation} 
\hat{M} = \frac{1}{4}(c_1+c_2)(X_1 X_2 + Y_1 Y_2) + \frac{i}{4}(c_1-c_2)(X_1Y_2-Y_1X_2) = \begin{pmatrix}
0 & 0 & 0 & 0 \\ 0 & 0 & c_2 & 0 \\ 0 & c_1 & 0  & 0 \\ 0 & 0 & 0 & 0
\end{pmatrix}.
\end{equation}
One of the operators in the Pauli basis, $A_{XX}$, can be projected as follows:
\begin{equation}
\hat{A}_{XX}^c = \sum_i P_i (X_1 X_2) P_i = \begin{pmatrix} 0 & 0 & 0 & 0 \\ 0 & 0 & 1 & 0  \\
0 & 1 & 0 & 0 \\ 0 & 0 & 0 & 0
\end{pmatrix}.~~
\end{equation}
To express this as a vector in the computational basis in the operator space, let $e_{i,j}$ denote a basis element where $i,j$ denote the row and column index, which gives:
\begin{equation}
\vec{A}_{XX} = e_{10,01} + e_{01,10}.
\end{equation} 
Similar vector forms can be found for the other Pauli matrices:
\begin{align}
\vec{A}_{XY}^c  &= -i e_{10,01}+ i e_{01,10}\\ 
\vec{A}_{YX}^c &= i e_{10,01}- i e_{01,10}\\ 
\vec{A}_{YY}^c  &= e_{10,01} + e_{01,10}. 
\end{align}
Clearly, we are limited by the dimension of the span of these vectors, leaving us to choose two vectors. Using $\vec{A}^c_{XX}$ and $\vec{A}^c_{XY}$ as the column vectors of $U$, and $\vec{m}$ as the target vector, we can find the solution to the system of linear equations $U\vec{x} = \vec{m}$:
\begin{equation}
\begin{pmatrix}
1 & -i  \\ 1 & i
\end{pmatrix}\begin{pmatrix} x_1  \\ x_2 \
\end{pmatrix}= \begin{pmatrix}
c_1 \\ c_2 
\end{pmatrix} \rightarrow \vec{x} = \frac{1}{2}\begin{pmatrix}
c_1 + c_2 \\ ic_1 - ic_2 
\end{pmatrix}.
\end{equation}
Thus a measurement of the form:
\begin{equation}
\hat{X} = \frac{c_1+c_2}{2} X_1 X_2 	+ \frac{i(c_1-c_2)}{2} X_1 Y_2 
\end{equation}
yields equivalent information as the traditional measurement due to symmetries of the system. 
}
\textcolor{black}{
\section{Computational Details}
Figure 1 was created by first generating the set of 2-RDM operators for a given molecular system and transforming them into a set of Pauli operators using fermionic-to-qubit transformations followed by our symmetry projection technique. To efficiently group terms, we expressed the set of operators as a graph utilizing the qubit-wise commuting relationship, and used an algorithm to attempt to find the minimimum clique cover \cite{Verteletskyi2019}. The graphs were stored using the graph-tools (v 2.29) Python package \cite{peixoto_graph-tool_2014}. A sequential coloring algorithm was used where we selected the vertices with the largest number of edges first, as this proved to be a reliable approach which was scalable to larger qubit systems which did not yield significantly worse results than the related recursive method \cite{Verteletskyi2019}. 
}

\textcolor{black}{
Table 3 was generated with the help of Qiskit (v 0.15.0) \cite{Qiskit}, and simulates a two-electron system in a minimal basis (STO-3G) under the Jordan-Wigner transformation on four qubits. We prepared 25 random states parameterized by a double excitation and two single excitations. The experimental results were obtained on the IBMQ Bogota device. The noise model was based on the backend-centered model in the Aer module of Qiskit, which approximates the noise channels mainly as a product of depolarizing and thermal relaxation channels acting locally on single- and two-qubit gates. The parameters were based on averages of the experimental devices, and were scaled down according to $\tau$ to model a decrease in the strength of the noise. 
}
%

%
%
%
\section{Further Computational Details}
For the quantum computation and noise simulation in Table III of the main text, we used the quantum computer IBMQ Bogota (5-qubits) provided through the IBM Quantum Experience, as well as a noise model based on the device. The quantum device has fixed-frequency transmon qubits with co-planer waveguide resonators~\cite{Koch2007,Chow2011}. The Python package Qiskit (v 0.15.0) \cite{Qiskit} was used to interface with the device. Device properties can be found in Supplemental Table IV.

\renewcommand\tablename{\textbf{Supplemental Table}}
\begin{table}[h]\caption{Calibration data for the ibm-bogota device taken on November $2^{\text{nd}}$, 2020, from benchmarking. U$_2$ and U$_3$ represent single qubit gate errors containing one and two $X_{\pi /2}$ pulses and two and three frame changes respectively. RO$_{i|j}$ represents the probability of measuring the state $i$ given a prepared state $j$. $T_1$ and $T_2$ are the given thermal relaxation times for each qubit. Frequency refers to the qubits opeartional frequency, and influences the excited state population based on the device temperature. [$j$] specifies the target qubit with control qubit $i$, and the number in paranthesis after each entry in the CNOT column indicates the gate length. The gate lengths for the $U_2$ and $U_3$ gates were 35 ns and 71 ns respectively.}
\begin{tabular}{c|c|cc|cc|cc|cc}
\hline
\textbf{Qubit} & \textbf{Frequency} & \textbf{U}$_2$  & \textbf{U}$_3$  & \textbf{RO$_{0|1}$} & \textbf{RO$_{1|0}$}& \textbf{T}$_1$  & \textbf{T}$_2$  & \multicolumn{2}{c}{$\left[ j \right]$ \textbf{CNOT}$_i^j$ } \\
$i$ & GHz & ($10^{-4}$) & ($10^{-4}$) &  ($10^{-2}$)& ($10^{-2}$)  & ($\mu s$) & ($\mu s$)  & \multicolumn{2}{c}{($10^{-2}$)}  \\
\hline 
0  & 5.000& 3.7  & 4.5  & 3.6 & 8.0 & 93.6 & 133.3 & [1] 2.0 (690) &  \\
1  & 4.845&   3.2  & 6.5 &  17.3 & 15.1 & 59.9 & 58.5 & [0] 2.0 (654) & [2] 1.0 (498) \\
2  &4.783  & 1.7  & 3.3  & 5.7 & 3.6 & 77.7 & 120.6 & [1] 1.0  (533)   &  [3] 1.0 (626)   \\
3  &4.858 & 2.4  & 4.8  &  3.0 & 0.9 & 131.1 & 187.1 & [2] 1.0 (590) & [4] 4.8 (370)  \\
4  &4.978 & 13.8 & 27.5 & 5.2 & 2.6 & 101.7 &  & [3] 4.8 (334) &         \\
\hline 
\end{tabular}
\end{table}

The circuit used is based on the Jordan-Wigner transformation and uses exponentials of anti-Hermitian operators. Because of the size of the system and the indistinguishablity of the action of the different Pauli operators on the state, we can use a single term to describe each of the relevant excitations \cite{Bonet-Monroig2018, Smart2020}. The target circuit can be written as:
\begin{equation}
U = \big[\exp \theta_1 (a^\dagger_{\alpha 0} a^{}_{\alpha 1} a^\dagger_{\beta 0 } a^{}_{\beta 1} - a^\dagger_{\beta 1} a^{}_{\beta 0} a^\dagger_{\alpha 1} a^{}_{\alpha 0})  \big] \big[ \exp \theta_2 (a^\dagger_{\alpha 0} a^{}_{\alpha 1} -  a^\dagger_{\alpha 1} a^{}_{\alpha 0} )  \big] \big[ \exp (\theta_3 a^\dagger_{\alpha 0} a^{}_{\alpha 1} -  a^\dagger_{\alpha 1} a^{}_{\alpha 0} ) \big]
\end{equation}
which can be simulated with limited Pauli terms as: 
\begin{equation}
U' =  [\exp i \theta_1  Y_1 X_2 X_3 X_4 ][\exp i \theta_2  Y_1 X_2][\exp i \theta_3   Y_3 X_4 ],
\end{equation}
and then simplified according to normal procedures. Our resulting circuit had 8 CNOT gates and 9 single qubit gates prior to measurement. We performed tomography of both the real and imaginary elements of the 2-RDM despite having only a real wavefunction. The list of measurement circuits generated for the normal circuit (following the grouping procedure) is:
\begin{align*}
N = \{~ 
&Y_1X_2Z_3Z_4, X_1X_2Z_3Z_4, Y_1Y_2Z_3Z_4, X_1Y_2Z_3Z_4, Z_1Z_2Y_3X_4, \\ &Z_1Z_2X_3X_4, Z_1Z_2Y_3Y_4, Z_1Z_2X_3Y_4, Y_1X_2Y_3X_4, X_1X_2Y_3X_4, \\ 
&Y_1Y_2Y_3X_4, X_1Y_2Y_3X_4, Y_1X_2X_3X_4, X_1X_2X_3X_4, Y_1Y_2X_3X_4, \\
&X_1Y_2X_3X_4, Y_1X_2Y_3Y_4, X_1X_2Y_3Y_4, Y_1Y_2Y_3Y_4, X_1Y_2Y_3Y_4, \\
& Y_1 X_2X_3 Y_4, X_1 X_2 X_3 Y_4, Y_1Y_2X_3Y_4, X_1Y_2X_3Y_4, Z_1Z_2Z_3Z_4 \},
\end{align*}
whereas the set of reduced circuits (with the same grouping procedure) is given by:
\begin{align*}
R = \{~ &X_1X_2Z_3Z_4, Y_1X_2Z_3Z_4, Z_1Z_2X_3X_4, Z_1Z_2Y_3X_4, X_1 X_2 X_3 X_4 , \\
& X_1 X_2 Y_3 X_4, Y_1 X_2 X_3 X_4, Y_1 X_2 Y_3 X_4, Z_1 Z_2 Z_3 Z_4 
\}. 
\end{align*}

\section{Noise Model}
The noise model used in generating the results in Table III is adapted from the provided noise model in Qiskit \cite{Qiskit}, and consists of a depolarizing channel followed by a thermal relaxation channel on each gate, with a readout error applied at measurement. Information on the model is adapted from the documentation provided in Qiskit \cite{Qiskit}.  Wood et al. is also referenced in the documentation and contains useful information regarding error channel representations and transformations \cite{Wood2015}. A useful online discussion of the benefits (applicability to short $T_1$ processes such as on single qubit gates) and limitations (non-$T_1$ dominated behavior of CNOT gates, does not treat cross-talk errors, etc.) of the noise model is included in the references \cite{Cjwood2019}. 

In our adaptation of the model, we averaged over all qubits for many of the parameters to reduce inconsistencies across the simulated device, and then scaled these parameters to simulate a consistently decreasing noise. We found that using either our model, the given model, or a model based solely on depolarizing noise did not result in signficant differences in the Frobenius norms of the difference matrices between the obtained 2-RDMs. 
\subsection{Thermal Relaxation Channel}
The $T_1$ time describes the thermal relaxation of the qubit from the excited state to the ground state and the $T_2$ time describes the coherence time of the qubit. Their respective relaxation rates for a given gate length $T_g$ are given as $r_{T_1} = \exp \frac{-T_g}{T_1}$ and $r_{T_2} = \exp \frac{-T_g}{T_2}$. For $T_2<T_1$, $T_1$ relaxation becomes the main consideration and the channel is expressed as a mixture of reset operations (proejctive measurements to $|0\rangle$ or $|1\rangle$) and unitary errors. Consider a ground state population $n_0$ and an excited state population:
\begin{equation}
n_1 = (1+\exp \frac{2hf}{k_B T })^{-1}
\end{equation}
where $h$ is Planck's constant, $k_B$ is the Boltzmann constant, $T$ and $f$ are the qubit temperature and frequency, and $n_0+n_1=1$. The probability of a reset error occuring is defined as: 
\begin{equation}
p_{reset} = 1 - r_{T_1}. 
\end{equation}
Using the populations we can obtain the probability to reset to the $|0\rangle$ state $p_{r0}$, the probability to reset to the $|1\rangle$ state $p_{r1}$, and the probability to apply a $Z$ gate $p_z$ as:
 \begin{equation}
p_{r0} = n_0 p_{reset}, ~~p_{r1} = n_1 p_{reset},~~ p_z = \frac{1}{2}(1-p_{reset})(1-\frac{r_{T_2}}{r_{T_1}}).
\end{equation}
For the case where $T_2 >T_1$, the Choi-matrix representation is used. For a nosie channel $\mathcal{E}$, the Choi-matrix in the column representation $\mathcal{C}$ is defined by:
\begin{equation}
\mathcal{C} = \sum_{i,j} | i \rangle \langle j | \otimes \mathcal{E}(|i \rangle \langle j| ), 
\end{equation}
and the resulting action on the state can be determined as
\begin{equation}
\mathcal{E}(\rho) = {\rm Tr_1} \mathcal{C}( \rho^T \otimes \mathbb{I})
\end{equation} 
where we trace over the first system. The Choi matrix used in the model is:
\begin{equation}
\mathcal{C} = \begin{pmatrix}
1 - n_1 p_{reset} & 0 & 0 & r_{T_2} \\
0 & p_e p_{reset} & 0 & 0 \\
0 & 0 & n_0 p_{reset} & 0 \\
r_{T_2} & 0 & 0 & 1-n_0p_{reset}
\end{pmatrix}. 
\end{equation}
The Choi matrix is then converted to Kraus operators and then to Pauli gates to be implemented in the simulation. 

\subsection{Depolarizing Channel}
The depolarizing channel acting on $n$ qubits is given as \cite{Nielsen2010}: 
\begin{align}
\mathcal{E}_{depol} = (1-\lambda)I + \lambda D.
\end{align} 
where $I$ is the identity channel, $D$ is the completely depolarizing channel,  and $0 \leq \lambda \leq \frac{4^n}{4^n-1}$ (for $n$ qubits) indicates the relative strength.

If we consider the total fidelity as a function of the thermal relaxation and depolarizing channels, we have that:
\begin{align}
F &=  F( \mathcal{E}_{depol} \circ \mathcal{E}_{relax} )  \\
&= (1-\lambda) F( \mathcal{E}_{relax} )+ \lambda F(D \circ \mathcal{E}_{relax} ) \\
& = (1-\lambda) F(\mathcal{E}_{relax})+ \lambda F(D) \\
&= F(\mathcal{E}_{relax})- \lambda \frac{ F(\mathcal{E}_{relax})d-1}{d}
\end{align}
where $d$ is the dimension of the system and the average fidelity of the depolarizing channel is $1/d$. From this we can solve for $\lambda$:
\begin{align}
\lambda = d\frac{F(\mathcal{E}_{relax})-F}{F(\mathcal{E}_{relax})d - 1}.
\end{align}
The gate fidelity of these channels is given by:
\begin{align}
F_{avg}(\mathcal{E},U) &= \int d \psi \langle \psi| U^\dagger \mathcal{E} (|\psi \rangle \! \langle \psi|) U |\psi \rangle \\
&= \frac{ F_{\text{pro}}(\mathcal{E},U)d +1}{d+1} \\
&= \frac{{\rm Tr}(S^\dagger_U S^\dagger_{\mathcal{E}})+d}{d(d+1)}
\end{align}
where $F_{pro}$ indicates the process fidelity and $S$ represents the superator representation of a quantum channel. With all of these components, the model applies the depolarizing channel onto the one- or two-qubit gate followed by the thermal relaxation channel applied to each individual qubit. 
\subsection{Readout Error}
Finally, the readout errors on the devices are treated as single qubit errors. These modify the output based on the probability of reading one output given another. For instance, measuring the state for the qubit $q$ $|0\rangle$ will give $|1\rangle$ with a probability $p=\textbf{RO}_{1|0}(q)$, and will give $|0\rangle$ with a probability $1-p$.

\bibliography{msr_vf}

\begin{thebibliography}{61}%
\makeatletter
\providecommand \@ifxundefined [1]{%
 \@ifx{#1\undefined}
}%
\providecommand \@ifnum [1]{%
 \ifnum #1\expandafter \@firstoftwo
 \else \expandafter \@secondoftwo
 \fi
}%
\providecommand \@ifx [1]{%
 \ifx #1\expandafter \@firstoftwo
 \else \expandafter \@secondoftwo
 \fi
}%
\providecommand \natexlab [1]{#1}%
\providecommand \enquote  [1]{``#1''}%
\providecommand \bibnamefont  [1]{#1}%
\providecommand \bibfnamefont [1]{#1}%
\providecommand \citenamefont [1]{#1}%
\providecommand \href@noop [0]{\@secondoftwo}%
\providecommand \href [0]{\begingroup \@sanitize@url \@href}%
\providecommand \@href[1]{\@@startlink{#1}\@@href}%
\providecommand \@@href[1]{\endgroup#1\@@endlink}%
\providecommand \@sanitize@url [0]{\catcode `\\12\catcode `\$12\catcode
  `\&12\catcode `\#12\catcode `\^12\catcode `\_12\catcode `\%12\relax}%
\providecommand \@@startlink[1]{}%
\providecommand \@@endlink[0]{}%
\providecommand \url  [0]{\begingroup\@sanitize@url \@url }%
\providecommand \@url [1]{\endgroup\@href {#1}{\urlprefix }}%
\providecommand \urlprefix  [0]{URL }%
\providecommand \Eprint [0]{\href }%
\providecommand \doibase [0]{http://dx.doi.org/}%
\providecommand \selectlanguage [0]{\@gobble}%
\providecommand \bibinfo  [0]{\@secondoftwo}%
\providecommand \bibfield  [0]{\@secondoftwo}%
\providecommand \translation [1]{[#1]}%
\providecommand \BibitemOpen [0]{}%
\providecommand \bibitemStop [0]{}%
\providecommand \bibitemNoStop [0]{.\EOS\space}%
\providecommand \EOS [0]{\spacefactor3000\relax}%
\providecommand \BibitemShut  [1]{\csname bibitem#1\endcsname}%
\let\auto@bib@innerbib\@empty
\bibitem [{\citenamefont {Szabo}\ and\ \citenamefont
  {Ostlund}(1996)}]{Szabo1996}%
  \BibitemOpen
  \bibfield  {author} {\bibinfo {author} {\bibfnamefont {A.}~\bibnamefont
  {Szabo}}\ and\ \bibinfo {author} {\bibfnamefont {N.~S.}\ \bibnamefont
  {Ostlund}},\ }\href@noop {} {\emph {\bibinfo {title} {{Modern Quantum
  Chemistry: Introduction to Advanced Electronic Structure Theory}}}}\
  (\bibinfo  {publisher} {Dover Publications},\ \bibinfo {address} {New York},\
  \bibinfo {year} {1996})\BibitemShut {NoStop}%
\bibitem [{\citenamefont {Helgaker}\ \emph {et~al.}(2000)\citenamefont
  {Helgaker}, \citenamefont {J{\o}rgensen},\ and\ \citenamefont
  {Olsen}}]{Helgaker2000}%
  \BibitemOpen
  \bibfield  {author} {\bibinfo {author} {\bibfnamefont {Trygve}\ \bibnamefont
  {Helgaker}}, \bibinfo {author} {\bibfnamefont {Poul}\ \bibnamefont
  {J{\o}rgensen}}, \ and\ \bibinfo {author} {\bibfnamefont {Jeppe}\
  \bibnamefont {Olsen}},\ }\href {\doibase 10.1002/9781119019572} {\emph
  {\bibinfo {title} {Molecular Electronic-Structure Theory}}}\ (\bibinfo
  {publisher} {John Wiley {\&} Sons, Ltd},\ \bibinfo {address} {Chichester,
  UK},\ \bibinfo {year} {2000})\ p.\ \bibinfo {pages} {908}\BibitemShut
  {NoStop}%
\bibitem [{\citenamefont {Coleman}\ and\ \citenamefont
  {Yukalov}(2000)}]{Coleman2000}%
  \BibitemOpen
  \bibfield  {author} {\bibinfo {author} {\bibfnamefont {A.J.}\ \bibnamefont
  {Coleman}}\ and\ \bibinfo {author} {\bibfnamefont {V.I.}\ \bibnamefont
  {Yukalov}},\ }\href@noop {} {\emph {\bibinfo {title} {{Reduced Density
  Matrices: Coulson's Challenge}}}}\ (\bibinfo  {publisher} {Springer},\
  \bibinfo {address} {Berlin Heidelberg New York},\ \bibinfo {year}
  {2000})\BibitemShut {NoStop}%
\bibitem [{\citenamefont {Mazziotti}(2007)}]{Mazziotti2007}%
  \BibitemOpen
  \bibinfo {editor} {\bibfnamefont {David~A.}\ \bibnamefont {Mazziotti}},\
  ed.,\ \href {\doibase 10.1002/0470106603} {\emph {\bibinfo {title} {Advances
  in Chemical Physics}}},\ \bibinfo {series} {Advances in Chemical Physics},
  Vol.\ \bibinfo {volume} {134}\ (\bibinfo  {publisher} {John Wiley {\&} Sons,
  Inc.},\ \bibinfo {address} {Hoboken, NJ, USA},\ \bibinfo {year} {2007})\ p.\
  \bibinfo {pages} {574}\BibitemShut {NoStop}%
\bibitem [{\citenamefont {Zhao}\ \emph {et~al.}(2004)\citenamefont {Zhao},
  \citenamefont {Braams}, \citenamefont {Fukuda}, \citenamefont {Overton},\
  and\ \citenamefont {Percus}}]{Zhao2004}%
  \BibitemOpen
  \bibfield  {author} {\bibinfo {author} {\bibfnamefont {Zhengji}\ \bibnamefont
  {Zhao}}, \bibinfo {author} {\bibfnamefont {Bastiaan~J.}\ \bibnamefont
  {Braams}}, \bibinfo {author} {\bibfnamefont {Mituhiro}\ \bibnamefont
  {Fukuda}}, \bibinfo {author} {\bibfnamefont {Michael~L.}\ \bibnamefont
  {Overton}}, \ and\ \bibinfo {author} {\bibfnamefont {Jerome~K.}\ \bibnamefont
  {Percus}},\ }\bibfield  {title} {\enquote {\bibinfo {title} {The reduced
  density matrix method for electronic structure calculations and the role of
  three-index representability conditions},}\ }\href {\doibase
  10.1063/1.1636721} {\bibfield  {journal} {\bibinfo  {journal} {J. Chem.
  Phys.}\ }\textbf {\bibinfo {volume} {120}},\ \bibinfo {pages} {2095--2104}
  (\bibinfo {year} {2004})}\BibitemShut {NoStop}%
\bibitem [{\citenamefont {Mazziotti}(2004)}]{Mazziotti2004a}%
  \BibitemOpen
  \bibfield  {author} {\bibinfo {author} {\bibfnamefont {D.~A.}\ \bibnamefont
  {Mazziotti}},\ }\bibfield  {title} {\enquote {\bibinfo {title} {Realization
  of quantum chemistry without wave functions through first-order semidefinite
  programming},}\ }\href {\doibase 10.1103/PhysRevLett.93.213001} {\bibfield
  {journal} {\bibinfo  {journal} {Phys. Rev. Lett.}\ }\textbf {\bibinfo
  {volume} {93}},\ \bibinfo {pages} {213001} (\bibinfo {year}
  {2004})}\BibitemShut {NoStop}%
\bibitem [{\citenamefont {Shenvi}\ and\ \citenamefont
  {Izmaylov}(2010)}]{Shenvi2010}%
  \BibitemOpen
  \bibfield  {author} {\bibinfo {author} {\bibfnamefont {Neil}\ \bibnamefont
  {Shenvi}}\ and\ \bibinfo {author} {\bibfnamefont {Artur~F.}\ \bibnamefont
  {Izmaylov}},\ }\bibfield  {title} {\enquote {\bibinfo {title} {Active-{Space
  {N}}-representability constraints for variational two-particle reduced
  density matrix calculations},}\ }\href {\doibase
  10.1103/physrevlett.105.213003} {\bibfield  {journal} {\bibinfo  {journal}
  {Phys. Rev. Lett.}\ }\textbf {\bibinfo {volume} {105}},\ \bibinfo {pages}
  {213003} (\bibinfo {year} {2010})}\BibitemShut {NoStop}%
\bibitem [{\citenamefont {Mazziotti}(2011)}]{Mazziotti2011}%
  \BibitemOpen
  \bibfield  {author} {\bibinfo {author} {\bibfnamefont {D.~A.}\ \bibnamefont
  {Mazziotti}},\ }\bibfield  {title} {\enquote {\bibinfo {title} {Large-scale
  semidefinite programming for many-electron quantum mechanics},}\ }\href
  {\doibase 10.1103/PhysRevLett.106.083001} {\bibfield  {journal} {\bibinfo
  {journal} {Phys. Rev. Lett.}\ }\textbf {\bibinfo {volume} {106}},\ \bibinfo
  {pages} {083001} (\bibinfo {year} {2011})}\BibitemShut {NoStop}%
\bibitem [{\citenamefont {Verstichel}\ \emph {et~al.}(2012)\citenamefont
  {Verstichel}, \citenamefont {van Aggelen}, \citenamefont {Poelmans},\ and\
  \citenamefont {{Van Neck}}}]{Verstichel2012}%
  \BibitemOpen
  \bibfield  {author} {\bibinfo {author} {\bibfnamefont {Brecht}\ \bibnamefont
  {Verstichel}}, \bibinfo {author} {\bibfnamefont {Helen}\ \bibnamefont {van
  Aggelen}}, \bibinfo {author} {\bibfnamefont {Ward}\ \bibnamefont {Poelmans}},
  \ and\ \bibinfo {author} {\bibfnamefont {Dimitri}\ \bibnamefont {{Van
  Neck}}},\ }\bibfield  {title} {\enquote {\bibinfo {title} {Variational
  two-particle density matrix calculation for the hubbard model below half
  filling using spin-adapted lifting conditions},}\ }\href {\doibase
  10.1103/physrevlett.108.213001} {\bibfield  {journal} {\bibinfo  {journal}
  {Phys. Rev. Lett.}\ }\textbf {\bibinfo {volume} {108}},\ \bibinfo {pages}
  {213001} (\bibinfo {year} {2012})}\BibitemShut {NoStop}%
\bibitem [{\citenamefont {Schilling}\ \emph {et~al.}(2013)\citenamefont
  {Schilling}, \citenamefont {Gross},\ and\ \citenamefont
  {Christandl}}]{Schilling2013}%
  \BibitemOpen
  \bibfield  {author} {\bibinfo {author} {\bibfnamefont {C.}~\bibnamefont
  {Schilling}}, \bibinfo {author} {\bibfnamefont {D.}~\bibnamefont {Gross}}, \
  and\ \bibinfo {author} {\bibfnamefont {M.}~\bibnamefont {Christandl}},\
  }\bibfield  {title} {\enquote {\bibinfo {title} {Pinning of fermionic
  occupation numbers},}\ }\href {\doibase 10.1103/PhysRevLett.110.040404}
  {\bibfield  {journal} {\bibinfo  {journal} {Phys. Rev. Lett.}\ }\textbf
  {\bibinfo {volume} {110}},\ \bibinfo {pages} {040404} (\bibinfo {year}
  {2013})}\BibitemShut {NoStop}%
\bibitem [{\citenamefont {Mazziotti}(2016)}]{Mazziotti2016}%
  \BibitemOpen
  \bibfield  {author} {\bibinfo {author} {\bibfnamefont {D.~A.}\ \bibnamefont
  {Mazziotti}},\ }\bibfield  {title} {\enquote {\bibinfo {title} {Enhanced
  constraints for accurate lower bounds on many-electron quantum energies from
  variational two-electron reduced density matrix theory},}\ }\href {\doibase
  10.1103/PhysRevLett.117.153001} {\bibfield  {journal} {\bibinfo  {journal}
  {Phys. Rev. Lett.}\ }\textbf {\bibinfo {volume} {117}},\ \bibinfo {pages}
  {153001} (\bibinfo {year} {2016})}\BibitemShut {NoStop}%
\bibitem [{\citenamefont {Piris}(2017)}]{Piris2017a}%
  \BibitemOpen
  \bibfield  {author} {\bibinfo {author} {\bibfnamefont {M.}~\bibnamefont
  {Piris}},\ }\bibfield  {title} {\enquote {\bibinfo {title} {Global method for
  electron correlation},}\ }\href {\doibase 10.1103/PhysRevLett.119.063002}
  {\bibfield  {journal} {\bibinfo  {journal} {Phys. Rev. Lett.}\ }\textbf
  {\bibinfo {volume} {119}},\ \bibinfo {pages} {063002} (\bibinfo {year}
  {2017})}\BibitemShut {NoStop}%
\bibitem [{\citenamefont {Rubio-Garc{\'{\i}}a}\ \emph
  {et~al.}(2019)\citenamefont {Rubio-Garc{\'{\i}}a}, \citenamefont {Dukelsky},
  \citenamefont {Alcoba}, \citenamefont {Capuzzi}, \citenamefont {O{\~{n}}a},
  \citenamefont {R{\'{\i}}os}, \citenamefont {Torre},\ and\ \citenamefont
  {Lain}}]{Rubio-Garcia2019}%
  \BibitemOpen
  \bibfield  {author} {\bibinfo {author} {\bibfnamefont {A.}~\bibnamefont
  {Rubio-Garc{\'{\i}}a}}, \bibinfo {author} {\bibfnamefont {J.}~\bibnamefont
  {Dukelsky}}, \bibinfo {author} {\bibfnamefont {D.~R.}\ \bibnamefont
  {Alcoba}}, \bibinfo {author} {\bibfnamefont {P.}~\bibnamefont {Capuzzi}},
  \bibinfo {author} {\bibfnamefont {O.~B.}\ \bibnamefont {O{\~{n}}a}}, \bibinfo
  {author} {\bibfnamefont {E.}~\bibnamefont {R{\'{\i}}os}}, \bibinfo {author}
  {\bibfnamefont {A.}~\bibnamefont {Torre}}, \ and\ \bibinfo {author}
  {\bibfnamefont {L.}~\bibnamefont {Lain}},\ }\bibfield  {title} {\enquote
  {\bibinfo {title} {Variational reduced density matrix method in the
  doubly-occupied configuration interaction space using
  four-{particleN}-representability conditions: Application to the {XXZ} model
  of quantum magnetism},}\ }\href {\doibase 10.1063/1.5118899} {\bibfield
  {journal} {\bibinfo  {journal} {J. Chem. Phys.}\ }\textbf {\bibinfo {volume}
  {151}},\ \bibinfo {pages} {154104} (\bibinfo {year} {2019})}\BibitemShut
  {NoStop}%
\bibitem [{\citenamefont {Mazziotti}(2006)}]{Mazziotti2006}%
  \BibitemOpen
  \bibfield  {author} {\bibinfo {author} {\bibfnamefont {D.~A.}\ \bibnamefont
  {Mazziotti}},\ }\bibfield  {title} {\enquote {\bibinfo {title}
  {Anti-{H}ermitian contracted {S}chr{\"o}dinger equation: Direct determination
  of the two-electron reduced density matrices of many-electron molecules},}\
  }\href {\doibase 10.1103/PhysRevLett.97.143002} {\bibfield  {journal}
  {\bibinfo  {journal} {Phys. Rev. Lett.}\ }\textbf {\bibinfo {volume} {97}},\
  \bibinfo {pages} {143002} (\bibinfo {year} {2006})}\BibitemShut {NoStop}%
\bibitem [{\citenamefont {Smart}\ and\ \citenamefont
  {Mazziotti}()}]{Smart2020a}%
  \BibitemOpen
  \bibfield  {author} {\bibinfo {author} {\bibfnamefont {S.~E.}\ \bibnamefont
  {Smart}}\ and\ \bibinfo {author} {\bibfnamefont {D.~A.}\ \bibnamefont
  {Mazziotti}},\ }\bibfield  {title} {\enquote {\bibinfo {title} {Quantum
  solver of contracted eigenvalue equations for scalable molecular simulations
  on quantum computing devices},}\ }\href@noop {} {\ }\Eprint
  {http://arxiv.org/abs/http://arxiv.org/abs/2004.11416v1}
  {http://arxiv.org/abs/2004.11416v1} \BibitemShut {NoStop}%
\bibitem [{\citenamefont {Coleman}(1963)}]{Coleman1963}%
  \BibitemOpen
  \bibfield  {author} {\bibinfo {author} {\bibfnamefont {A.~J.}\ \bibnamefont
  {Coleman}},\ }\bibfield  {title} {\enquote {\bibinfo {title} {Structure of
  fermion density matrices},}\ }\href {\doibase 10.1103/revmodphys.35.668}
  {\bibfield  {journal} {\bibinfo  {journal} {Rev. Mod. Phys.}\ }\textbf
  {\bibinfo {volume} {35}},\ \bibinfo {pages} {668--686} (\bibinfo {year}
  {1963})}\BibitemShut {NoStop}%
\bibitem [{\citenamefont {Mazziotti}(2012)}]{Mazziotti2012}%
  \BibitemOpen
  \bibfield  {author} {\bibinfo {author} {\bibfnamefont {David~A.}\
  \bibnamefont {Mazziotti}},\ }\bibfield  {title} {\enquote {\bibinfo {title}
  {{Structure of Fermionic Density Matrices: Complete N-Representability
  Conditions}},}\ }\href {\doibase 10.1103/PhysRevLett.108.263002} {\bibfield
  {journal} {\bibinfo  {journal} {Phys. Rev. Lett.}\ }\textbf {\bibinfo
  {volume} {108}},\ \bibinfo {pages} {263002} (\bibinfo {year} {2012})},\
  \Eprint {http://arxiv.org/abs/1112.5866} {arXiv:1112.5866} \BibitemShut
  {NoStop}%
\bibitem [{\citenamefont {Boyn}\ \emph {et~al.}(2020)\citenamefont {Boyn},
  \citenamefont {Xie}, \citenamefont {Anderson},\ and\ \citenamefont
  {Mazziotti}}]{Boyn2020}%
  \BibitemOpen
  \bibfield  {author} {\bibinfo {author} {\bibfnamefont {Jan-Niklas}\
  \bibnamefont {Boyn}}, \bibinfo {author} {\bibfnamefont {Jiaze}\ \bibnamefont
  {Xie}}, \bibinfo {author} {\bibfnamefont {John~S.}\ \bibnamefont {Anderson}},
  \ and\ \bibinfo {author} {\bibfnamefont {David~A.}\ \bibnamefont
  {Mazziotti}},\ }\bibfield  {title} {\enquote {\bibinfo {title} {Entangled
  electrons drive a non-superexchange mechanism in a cobalt quinoid dimer
  complex},}\ }\href {\doibase 10.1021/acs.jpclett.0c01248} {\bibfield
  {journal} {\bibinfo  {journal} {J. Phys. Chem. Lett.}\ ,\ \bibinfo {pages}
  {4584--4590}} (\bibinfo {year} {2020})}\BibitemShut {NoStop}%
\bibitem [{\citenamefont {Feynman}\ \emph {et~al.}(1963--1965)\citenamefont
  {Feynman}, \citenamefont {Leighton},\ and\ \citenamefont
  {Sands}}]{Feynman1963}%
  \BibitemOpen
  \bibfield  {author} {\bibinfo {author} {\bibfnamefont {Richard P.
  (Richard~Phillips)}\ \bibnamefont {Feynman}}, \bibinfo {author}
  {\bibfnamefont {Robert~B.}\ \bibnamefont {Leighton}}, \ and\ \bibinfo
  {author} {\bibfnamefont {Matthew L. (Matthew~Linzee)}\ \bibnamefont
  {Sands}},\ }\href@noop {} {\emph {\bibinfo {title} {The {Feynman} lectures on
  physics}}}\ (\bibinfo {year} {1963--1965})\ pp.\ \bibinfo {pages} {xii +
  513},\ \bibinfo {note} {three volumes.}\BibitemShut {Stop}%
\bibitem [{\citenamefont {Bishop}(1993)}]{Bishop1993}%
  \BibitemOpen
  \bibfield  {author} {\bibinfo {author} {\bibfnamefont {David~M.}\
  \bibnamefont {Bishop}},\ }\href@noop {} {\emph {\bibinfo {title} {{Group
  Theory and Chemistry}}}}\ (\bibinfo  {publisher} {Dover Publications},\
  \bibinfo {address} {Mineola, N.Y.},\ \bibinfo {year} {1993})\BibitemShut
  {NoStop}%
\bibitem [{\citenamefont {Gidofalvi}\ and\ \citenamefont
  {Mazziotti}(2005)}]{Gidofalvi2005}%
  \BibitemOpen
  \bibfield  {author} {\bibinfo {author} {\bibfnamefont {Gergely}\ \bibnamefont
  {Gidofalvi}}\ and\ \bibinfo {author} {\bibfnamefont {David~A}\ \bibnamefont
  {Mazziotti}},\ }\bibfield  {title} {\enquote {\bibinfo {title} {{Spin and
  symmetry adaptation of the variational two-electron reduced-density-matrix
  method}},}\ }\href {\doibase 10.1103/PhysRevA.72.052505} {\bibfield
  {journal} {\bibinfo  {journal} {Phys. Rev. A}\ }\textbf {\bibinfo {volume}
  {72}},\ \bibinfo {pages} {052505} (\bibinfo {year} {2005})}\BibitemShut
  {NoStop}%
\bibitem [{\citenamefont {Nakatsuji}(1979)}]{Nakatsuji1979}%
  \BibitemOpen
  \bibfield  {author} {\bibinfo {author} {\bibfnamefont {Hiroshi}\ \bibnamefont
  {Nakatsuji}},\ }\bibfield  {title} {\enquote {\bibinfo {title} {{Cluster
  expansion of the wavefunction. Electron correlations in ground and excited
  states by SAC (symmetry-adapted-cluster) and SAC CI theories}},}\ }\href
  {\doibase 10.1016/0009-2614(79)85172-6} {\bibfield  {journal} {\bibinfo
  {journal} {Chemical Physics Letters}\ }\textbf {\bibinfo {volume} {67}},\
  \bibinfo {pages} {329--333} (\bibinfo {year} {1979})}\BibitemShut {NoStop}%
\bibitem [{\citenamefont {Rubin}\ \emph {et~al.}(2018)\citenamefont {Rubin},
  \citenamefont {Babbush},\ and\ \citenamefont {McClean}}]{Rubin2018}%
  \BibitemOpen
  \bibfield  {author} {\bibinfo {author} {\bibfnamefont {Nicholas~C.}\
  \bibnamefont {Rubin}}, \bibinfo {author} {\bibfnamefont {Ryan}\ \bibnamefont
  {Babbush}}, \ and\ \bibinfo {author} {\bibfnamefont {Jarrod}\ \bibnamefont
  {McClean}},\ }\bibfield  {title} {\enquote {\bibinfo {title} {{Application of
  fermionic marginal constraints to hybrid quantum algorithms}},}\ }\href
  {\doibase 10.1088/1367-2630/aab919} {\bibfield  {journal} {\bibinfo
  {journal} {New J. Phys.}\ }\textbf {\bibinfo {volume} {20}},\ \bibinfo
  {pages} {053020} (\bibinfo {year} {2018})},\ \Eprint
  {http://arxiv.org/abs/1801.03524} {arXiv:1801.03524} \BibitemShut {NoStop}%
\bibitem [{\citenamefont {Bonet-Monroig}\ \emph {et~al.}(2019)\citenamefont
  {Bonet-Monroig}, \citenamefont {Babbush},\ and\ \citenamefont
  {O'Brien}}]{Bonet-Monroig2019}%
  \BibitemOpen
  \bibfield  {author} {\bibinfo {author} {\bibfnamefont {Xavier}\ \bibnamefont
  {Bonet-Monroig}}, \bibinfo {author} {\bibfnamefont {Ryan}\ \bibnamefont
  {Babbush}}, \ and\ \bibinfo {author} {\bibfnamefont {Thomas~E}\ \bibnamefont
  {O'Brien}},\ }\bibfield  {title} {\enquote {\bibinfo {title} {{Nearly Optimal
  Measurement Scheduling for Partial Tomography of Quantum States}},}\ }\href
  {http://arxiv.org/abs/1908.05628} {\ ,\ \bibinfo {pages} {1--9} (\bibinfo
  {year} {2019})},\ \Eprint {http://arxiv.org/abs/1908.05628}
  {arXiv:1908.05628} \BibitemShut {NoStop}%
\bibitem [{\citenamefont {Smart}\ and\ \citenamefont
  {Mazziotti}(2019)}]{Smart2019}%
  \BibitemOpen
  \bibfield  {author} {\bibinfo {author} {\bibfnamefont {Scott~E}\ \bibnamefont
  {Smart}}\ and\ \bibinfo {author} {\bibfnamefont {David~A}\ \bibnamefont
  {Mazziotti}},\ }\bibfield  {title} {\enquote {\bibinfo {title}
  {{Quantum-classical hybrid algorithm using an error-mitigating
  N-representability condition to compute the Mott metal-insulator
  transition}},}\ }\href {\doibase 10.1103/PhysRevA.100.022517} {\bibfield
  {journal} {\bibinfo  {journal} {Phys. Rev. A}\ }\textbf {\bibinfo {volume}
  {100}},\ \bibinfo {pages} {022517} (\bibinfo {year} {2019})}\BibitemShut
  {NoStop}%
\bibitem [{\citenamefont {Sager}\ \emph {et~al.}()\citenamefont {Sager},
  \citenamefont {Smart},\ and\ \citenamefont {Mazziotti}}]{Sager2020}%
  \BibitemOpen
  \bibfield  {author} {\bibinfo {author} {\bibfnamefont {LeeAnn~M.}\
  \bibnamefont {Sager}}, \bibinfo {author} {\bibfnamefont {Scott~E.}\
  \bibnamefont {Smart}}, \ and\ \bibinfo {author} {\bibfnamefont {David~A.}\
  \bibnamefont {Mazziotti}},\ }\bibfield  {title} {\enquote {\bibinfo {title}
  {Preparation of an exciton condensate of photons on a 53-qubit quantum
  computer},}\ }\href@noop {} {\ }\Eprint
  {http://arxiv.org/abs/http://arxiv.org/abs/2004.13868v1}
  {http://arxiv.org/abs/2004.13868v1} \BibitemShut {NoStop}%
\bibitem [{\citenamefont {Roth}\ \emph {et~al.}(2017)\citenamefont {Roth},
  \citenamefont {Ganzhorn}, \citenamefont {Moll}, \citenamefont {Filipp},
  \citenamefont {Salis},\ and\ \citenamefont {Schmidt}}]{Roth2017}%
  \BibitemOpen
  \bibfield  {author} {\bibinfo {author} {\bibfnamefont {Marco}\ \bibnamefont
  {Roth}}, \bibinfo {author} {\bibfnamefont {Marc}\ \bibnamefont {Ganzhorn}},
  \bibinfo {author} {\bibfnamefont {Nikolaj}\ \bibnamefont {Moll}}, \bibinfo
  {author} {\bibfnamefont {Stefan}\ \bibnamefont {Filipp}}, \bibinfo {author}
  {\bibfnamefont {Gian}\ \bibnamefont {Salis}}, \ and\ \bibinfo {author}
  {\bibfnamefont {Sebastian}\ \bibnamefont {Schmidt}},\ }\bibfield  {title}
  {\enquote {\bibinfo {title} {{Analysis of a parametrically driven
  exchange-type gate and a two-photon excitation gate between superconducting
  qubits}},}\ }\href {\doibase 10.1103/PhysRevA.96.062323} {\bibfield
  {journal} {\bibinfo  {journal} {Physical Review A}\ }\textbf {\bibinfo
  {volume} {96}},\ \bibinfo {pages} {1--10} (\bibinfo {year} {2017})},\ \Eprint
  {http://arxiv.org/abs/arXiv:1708.02090v1} {arXiv:arXiv:1708.02090v1}
  \BibitemShut {NoStop}%
\bibitem [{\citenamefont {Ganzhorn}\ \emph {et~al.}(2019)\citenamefont
  {Ganzhorn}, \citenamefont {Egger}, \citenamefont {Ollitrault}, \citenamefont
  {Salis}, \citenamefont {Moll}, \citenamefont {Roth}, \citenamefont {Fuhrer},
  \citenamefont {Mueller}, \citenamefont {Woerner}, \citenamefont
  {Tavernelli},\ and\ \citenamefont {Filipp}}]{Ganzhorn2019}%
  \BibitemOpen
  \bibfield  {author} {\bibinfo {author} {\bibfnamefont {M.}~\bibnamefont
  {Ganzhorn}}, \bibinfo {author} {\bibfnamefont {P.}~\bibnamefont {Egger},
  \bibfnamefont {D.~J. an d~Barkoutsos}}, \bibinfo {author} {\bibfnamefont
  {P.}~\bibnamefont {Ollitrault}}, \bibinfo {author} {\bibfnamefont
  {G.}~\bibnamefont {Salis}}, \bibinfo {author} {\bibfnamefont
  {N.}~\bibnamefont {Moll}}, \bibinfo {author} {\bibfnamefont {M.}~\bibnamefont
  {Roth}}, \bibinfo {author} {\bibfnamefont {A.}~\bibnamefont {Fuhrer}},
  \bibinfo {author} {\bibfnamefont {P.}~\bibnamefont {Mueller}}, \bibinfo
  {author} {\bibfnamefont {S.}~\bibnamefont {Woerner}}, \bibinfo {author}
  {\bibfnamefont {I.}~\bibnamefont {Tavernelli}}, \ and\ \bibinfo {author}
  {\bibfnamefont {S.}~\bibnamefont {Filipp}},\ }\bibfield  {title} {\enquote
  {\bibinfo {title} {{Gate-Efficient Simulation of Molecular Eigenstates on a
  Quantum Computer}},}\ }\href {\doibase 10.1103/PhysRevApplied.11.044092}
  {\bibfield  {journal} {\bibinfo  {journal} {Physical Review Applied}\
  }\textbf {\bibinfo {volume} {11}},\ \bibinfo {pages} {1} (\bibinfo {year}
  {2019})},\ \Eprint {http://arxiv.org/abs/1809.05057} {arXiv:1809.05057}
  \BibitemShut {NoStop}%
\bibitem [{\citenamefont {Wang}\ \emph {et~al.}(2009)\citenamefont {Wang},
  \citenamefont {Ashhab},\ and\ \citenamefont {Nori}}]{Wang2009}%
  \BibitemOpen
  \bibfield  {author} {\bibinfo {author} {\bibfnamefont {Hefeng}\ \bibnamefont
  {Wang}}, \bibinfo {author} {\bibfnamefont {S.}~\bibnamefont {Ashhab}}, \ and\
  \bibinfo {author} {\bibfnamefont {Franco}\ \bibnamefont {Nori}},\ }\bibfield
  {title} {\enquote {\bibinfo {title} {{Efficient quantum algorithm for
  preparing molecular-system-like states on a quantum computer}},}\ }\href
  {\doibase 10.1103/PhysRevA.79.042335} {\bibfield  {journal} {\bibinfo
  {journal} {Physical Review A - Atomic, Molecular, and Optical Physics}\
  }\textbf {\bibinfo {volume} {79}} (\bibinfo {year} {2009}),\
  10.1103/PhysRevA.79.042335},\ \Eprint {http://arxiv.org/abs/0902.1419}
  {arXiv:0902.1419} \BibitemShut {NoStop}%
\bibitem [{\citenamefont {Whitfield}(2013)}]{Whitfield2013}%
  \BibitemOpen
  \bibfield  {author} {\bibinfo {author} {\bibfnamefont {James~Daniel}\
  \bibnamefont {Whitfield}},\ }\bibfield  {title} {\enquote {\bibinfo {title}
  {{Communication: Spin-free quantum computational simulations and symmetry
  adapted states}},}\ }\href {\doibase 10.1063/1.4812566} {\bibfield  {journal}
  {\bibinfo  {journal} {Journal of Chemical Physics}\ }\textbf {\bibinfo
  {volume} {139}} (\bibinfo {year} {2013}),\ 10.1063/1.4812566}\BibitemShut
  {NoStop}%
\bibitem [{\citenamefont {Barkoutsos}\ \emph {et~al.}(2018)\citenamefont
  {Barkoutsos}, \citenamefont {Gonthier}, \citenamefont {Sokolov},
  \citenamefont {Moll}, \citenamefont {Salis}, \citenamefont {Fuhrer},
  \citenamefont {Ganzhorn}, \citenamefont {Egger}, \citenamefont {Troyer},
  \citenamefont {Mezzacapo}, \citenamefont {Filipp},\ and\ \citenamefont
  {Tavernelli}}]{Barkoutsos2018}%
  \BibitemOpen
  \bibfield  {author} {\bibinfo {author} {\bibfnamefont {Panagiotis~Kl}\
  \bibnamefont {Barkoutsos}}, \bibinfo {author} {\bibfnamefont {Jerome~F.}\
  \bibnamefont {Gonthier}}, \bibinfo {author} {\bibfnamefont {Igor}\
  \bibnamefont {Sokolov}}, \bibinfo {author} {\bibfnamefont {Nikolaj}\
  \bibnamefont {Moll}}, \bibinfo {author} {\bibfnamefont {Gian}\ \bibnamefont
  {Salis}}, \bibinfo {author} {\bibfnamefont {Andreas}\ \bibnamefont {Fuhrer}},
  \bibinfo {author} {\bibfnamefont {Marc}\ \bibnamefont {Ganzhorn}}, \bibinfo
  {author} {\bibfnamefont {Daniel~J.}\ \bibnamefont {Egger}}, \bibinfo {author}
  {\bibfnamefont {Matthias}\ \bibnamefont {Troyer}}, \bibinfo {author}
  {\bibfnamefont {Antonio}\ \bibnamefont {Mezzacapo}}, \bibinfo {author}
  {\bibfnamefont {Stefan}\ \bibnamefont {Filipp}}, \ and\ \bibinfo {author}
  {\bibfnamefont {Ivano}\ \bibnamefont {Tavernelli}},\ }\bibfield  {title}
  {\enquote {\bibinfo {title} {{Quantum algorithms for electronic structure
  calculations: Particle-hole Hamiltonian and optimized wave-function
  expansions}},}\ }\href {\doibase 10.1103/PhysRevA.98.022322} {\bibfield
  {journal} {\bibinfo  {journal} {Physical Review A}\ }\textbf {\bibinfo
  {volume} {98}},\ \bibinfo {pages} {022322} (\bibinfo {year} {2018})},\
  \Eprint {http://arxiv.org/abs/1805.04340} {arXiv:1805.04340} \BibitemShut
  {NoStop}%
\bibitem [{\citenamefont {Fischer}\ and\ \citenamefont
  {Gunlycke}(2019)}]{Fischer2019}%
  \BibitemOpen
  \bibfield  {author} {\bibinfo {author} {\bibfnamefont {Sean~A.}\ \bibnamefont
  {Fischer}}\ and\ \bibinfo {author} {\bibfnamefont {Daniel}\ \bibnamefont
  {Gunlycke}},\ }\bibfield  {title} {\enquote {\bibinfo {title} {{Symmetry
  Configuration Mapping for Representing Quantum Systems on Quantum
  Computers}},}\ }\href {http://arxiv.org/abs/1907.01493} {\  (\bibinfo {year}
  {2019})},\ \Eprint {http://arxiv.org/abs/1907.01493} {arXiv:1907.01493}
  \BibitemShut {NoStop}%
\bibitem [{\citenamefont {Barron}\ \emph {et~al.}(2020)\citenamefont {Barron},
  \citenamefont {Gard}, \citenamefont {Altman}, \citenamefont {Mayhall},
  \citenamefont {Barnes}, \citenamefont {Economou},\ and\ \citenamefont
  {Tech}}]{Barron2020}%
  \BibitemOpen
  \bibfield  {author} {\bibinfo {author} {\bibfnamefont {George~S}\
  \bibnamefont {Barron}}, \bibinfo {author} {\bibfnamefont {Bryan~T}\
  \bibnamefont {Gard}}, \bibinfo {author} {\bibfnamefont {Orien~J}\
  \bibnamefont {Altman}}, \bibinfo {author} {\bibfnamefont {Nicholas~J}\
  \bibnamefont {Mayhall}}, \bibinfo {author} {\bibfnamefont {Edwin}\
  \bibnamefont {Barnes}}, \bibinfo {author} {\bibfnamefont {Sophia~E}\
  \bibnamefont {Economou}}, \ and\ \bibinfo {author} {\bibfnamefont {Virginia}\
  \bibnamefont {Tech}},\ }\bibfield  {title} {\enquote {\bibinfo {title}
  {{Preserving Symmetries for Variational Quantum Eigensolvers in the Presence
  of Noise}},}\ }\href@noop {} {\ ,\ \bibinfo {pages} {1--13} (\bibinfo {year}
  {2020})},\ \Eprint {http://arxiv.org/abs/arXiv:2003.00171v1}
  {arXiv:arXiv:2003.00171v1} \BibitemShut {NoStop}%
\bibitem [{\citenamefont {Gard}\ \emph {et~al.}(2020)\citenamefont {Gard},
  \citenamefont {Zhu}, \citenamefont {Barron}, \citenamefont {Mayhall},
  \citenamefont {Economou},\ and\ \citenamefont {Barnes}}]{Gard2020}%
  \BibitemOpen
  \bibfield  {author} {\bibinfo {author} {\bibfnamefont {Bryan~T}\ \bibnamefont
  {Gard}}, \bibinfo {author} {\bibfnamefont {Linghua}\ \bibnamefont {Zhu}},
  \bibinfo {author} {\bibfnamefont {George~S}\ \bibnamefont {Barron}}, \bibinfo
  {author} {\bibfnamefont {Nicholas~J}\ \bibnamefont {Mayhall}}, \bibinfo
  {author} {\bibfnamefont {Sophia~E.}\ \bibnamefont {Economou}}, \ and\
  \bibinfo {author} {\bibfnamefont {Edwin}\ \bibnamefont {Barnes}},\ }\bibfield
   {title} {\enquote {\bibinfo {title} {{Efficient symmetry-preserving state
  preparation circuits for the variational quantum eigensolver algorithm}},}\
  }\href {\doibase 10.1038/s41534-019-0240-1} {\bibfield  {journal} {\bibinfo
  {journal} {npj Quantum Inf.}\ }\textbf {\bibinfo {volume} {6}} (\bibinfo
  {year} {2020}),\ 10.1038/s41534-019-0240-1},\ \Eprint
  {http://arxiv.org/abs/1904.10910} {arXiv:1904.10910} \BibitemShut {NoStop}%
\bibitem [{\citenamefont {Smart}\ and\ \citenamefont
  {Mazziotti}(2020)}]{Smart2020}%
  \BibitemOpen
  \bibfield  {author} {\bibinfo {author} {\bibfnamefont {Scott~E.}\
  \bibnamefont {Smart}}\ and\ \bibinfo {author} {\bibfnamefont {David~A.}\
  \bibnamefont {Mazziotti}},\ }\bibfield  {title} {\enquote {\bibinfo {title}
  {Efficient two-electron ansatz for benchmarking quantum chemistry on a
  quantum computer},}\ }\href {\doibase 10.1103/physrevresearch.2.023048}
  {\bibfield  {journal} {\bibinfo  {journal} {Phys. Rev. Res.}\ }\textbf
  {\bibinfo {volume} {2}},\ \bibinfo {pages} {023048} (\bibinfo {year}
  {2020})}\BibitemShut {NoStop}%
\bibitem [{\citenamefont {McClean}\ \emph
  {et~al.}(2016{\natexlab{a}})\citenamefont {McClean}, \citenamefont {Romero},
  \citenamefont {Babbush},\ and\ \citenamefont {Aspuru-Guzik}}]{McClean2016}%
  \BibitemOpen
  \bibfield  {author} {\bibinfo {author} {\bibfnamefont {Jarrod~R.}\
  \bibnamefont {McClean}}, \bibinfo {author} {\bibfnamefont {Jonathan}\
  \bibnamefont {Romero}}, \bibinfo {author} {\bibfnamefont {Ryan}\ \bibnamefont
  {Babbush}}, \ and\ \bibinfo {author} {\bibfnamefont {Al{\'{a}}n}\
  \bibnamefont {Aspuru-Guzik}},\ }\bibfield  {title} {\enquote {\bibinfo
  {title} {{The theory of variational hybrid quantum-classical algorithms}},}\
  }\href {\doibase 10.1088/1367-2630/18/2/023023} {\bibfield  {journal}
  {\bibinfo  {journal} {New Journal of Physics}\ }\textbf {\bibinfo {volume}
  {18}},\ \bibinfo {pages} {023023} (\bibinfo {year} {2016}{\natexlab{a}})},\
  \Eprint {http://arxiv.org/abs/1509.04279} {arXiv:1509.04279} \BibitemShut
  {NoStop}%
\bibitem [{\citenamefont {Kandala}\ \emph {et~al.}(2017)\citenamefont
  {Kandala}, \citenamefont {Mezzacapo}, \citenamefont {Temme}, \citenamefont
  {Takita}, \citenamefont {Brink}, \citenamefont {Chow},\ and\ \citenamefont
  {Gambetta}}]{Kandala2017}%
  \BibitemOpen
  \bibfield  {author} {\bibinfo {author} {\bibfnamefont {Abhinav}\ \bibnamefont
  {Kandala}}, \bibinfo {author} {\bibfnamefont {Antonio}\ \bibnamefont
  {Mezzacapo}}, \bibinfo {author} {\bibfnamefont {Kristan}\ \bibnamefont
  {Temme}}, \bibinfo {author} {\bibfnamefont {Maika}\ \bibnamefont {Takita}},
  \bibinfo {author} {\bibfnamefont {Markus}\ \bibnamefont {Brink}}, \bibinfo
  {author} {\bibfnamefont {Jerry~M.}\ \bibnamefont {Chow}}, \ and\ \bibinfo
  {author} {\bibfnamefont {Jay~M.}\ \bibnamefont {Gambetta}},\ }\bibfield
  {title} {\enquote {\bibinfo {title} {{Hardware-efficient variational quantum
  eigensolver for small molecules and quantum magnets}},}\ }\href {\doibase
  10.1038/nature23879} {\bibfield  {journal} {\bibinfo  {journal} {Nature}\
  }\textbf {\bibinfo {volume} {549}},\ \bibinfo {pages} {242--246} (\bibinfo
  {year} {2017})},\ \Eprint {http://arxiv.org/abs/1704.05018}
  {arXiv:1704.05018} \BibitemShut {NoStop}%
\bibitem [{\citenamefont {Ryabinkin}\ \emph {et~al.}(2019)\citenamefont
  {Ryabinkin}, \citenamefont {Genin},\ and\ \citenamefont
  {Izmaylov}}]{Ryabinkin2019}%
  \BibitemOpen
  \bibfield  {author} {\bibinfo {author} {\bibfnamefont {Ilya~G.}\ \bibnamefont
  {Ryabinkin}}, \bibinfo {author} {\bibfnamefont {Scott~N.}\ \bibnamefont
  {Genin}}, \ and\ \bibinfo {author} {\bibfnamefont {Artur~F.}\ \bibnamefont
  {Izmaylov}},\ }\bibfield  {title} {\enquote {\bibinfo {title} {{Constrained
  Variational Quantum Eigensolver: Quantum Computer Search Engine in the Fock
  Space}},}\ }\href {\doibase 10.1021/acs.jctc.8b00943} {\bibfield  {journal}
  {\bibinfo  {journal} {Journal of Chemical Theory and Computation}\ }\textbf
  {\bibinfo {volume} {15}},\ \bibinfo {pages} {249--255} (\bibinfo {year}
  {2019})},\ \Eprint {http://arxiv.org/abs/1806.00461} {arXiv:1806.00461}
  \BibitemShut {NoStop}%
\bibitem [{\citenamefont {Moll}\ \emph {et~al.}(2016)\citenamefont {Moll},
  \citenamefont {Fuhrer}, \citenamefont {Staar},\ and\ \citenamefont
  {Tavernelli}}]{Moll2016}%
  \BibitemOpen
  \bibfield  {author} {\bibinfo {author} {\bibfnamefont {Nikolaj}\ \bibnamefont
  {Moll}}, \bibinfo {author} {\bibfnamefont {Andreas}\ \bibnamefont {Fuhrer}},
  \bibinfo {author} {\bibfnamefont {Peter}\ \bibnamefont {Staar}}, \ and\
  \bibinfo {author} {\bibfnamefont {Ivano}\ \bibnamefont {Tavernelli}},\
  }\bibfield  {title} {\enquote {\bibinfo {title} {{Optimizing qubit resources
  for quantum chemistry simulations in second quantization on a quantum
  computer}},}\ }\href {\doibase 10.1088/1751-8113/49/29/295301} {\bibfield
  {journal} {\bibinfo  {journal} {J. Phys. A: Math. Theor.}\ }\textbf {\bibinfo
  {volume} {49}} (\bibinfo {year} {2016}),\ 10.1088/1751-8113/49/29/295301},\
  \Eprint {http://arxiv.org/abs/1510.04048} {arXiv:1510.04048} \BibitemShut
  {NoStop}%
\bibitem [{\citenamefont {Bravyi}\ \emph {et~al.}(2017)\citenamefont {Bravyi},
  \citenamefont {Gambetta}, \citenamefont {Mezzacapo},\ and\ \citenamefont
  {Temme}}]{Bravyi2017}%
  \BibitemOpen
  \bibfield  {author} {\bibinfo {author} {\bibfnamefont {Sergey}\ \bibnamefont
  {Bravyi}}, \bibinfo {author} {\bibfnamefont {Jay~M.}\ \bibnamefont
  {Gambetta}}, \bibinfo {author} {\bibfnamefont {Antonio}\ \bibnamefont
  {Mezzacapo}}, \ and\ \bibinfo {author} {\bibfnamefont {Kristan}\ \bibnamefont
  {Temme}},\ }\bibfield  {title} {\enquote {\bibinfo {title} {{Tapering off
  qubits to simulate fermionic Hamiltonians}},}\ }\href
  {http://arxiv.org/abs/1701.08213} {\ ,\ \bibinfo {pages} {1--15} (\bibinfo
  {year} {2017})},\ \Eprint {http://arxiv.org/abs/1701.08213}
  {arXiv:1701.08213} \BibitemShut {NoStop}%
\bibitem [{\citenamefont {Setia}\ \emph {et~al.}(2019)\citenamefont {Setia},
  \citenamefont {Chen}, \citenamefont {Rice}, \citenamefont {Mezzacapo},
  \citenamefont {Pistoia},\ and\ \citenamefont {Whitfield}}]{Setia2019}%
  \BibitemOpen
  \bibfield  {author} {\bibinfo {author} {\bibfnamefont {Kanav}\ \bibnamefont
  {Setia}}, \bibinfo {author} {\bibfnamefont {Richard}\ \bibnamefont {Chen}},
  \bibinfo {author} {\bibfnamefont {Julia~E.}\ \bibnamefont {Rice}}, \bibinfo
  {author} {\bibfnamefont {Antonio}\ \bibnamefont {Mezzacapo}}, \bibinfo
  {author} {\bibfnamefont {Marco}\ \bibnamefont {Pistoia}}, \ and\ \bibinfo
  {author} {\bibfnamefont {James}\ \bibnamefont {Whitfield}},\ }\bibfield
  {title} {\enquote {\bibinfo {title} {{Reducing qubit requirements for quantum
  simulation using molecular point group symmetries}},}\ }\href
  {http://arxiv.org/abs/1910.14644} {\ ,\ \bibinfo {pages} {1--6} (\bibinfo
  {year} {2019})},\ \Eprint {http://arxiv.org/abs/1910.14644}
  {arXiv:1910.14644} \BibitemShut {NoStop}%
\bibitem [{\citenamefont {McArdle}\ \emph {et~al.}(2019)\citenamefont
  {McArdle}, \citenamefont {Yuan},\ and\ \citenamefont
  {Benjamin}}]{McArdle2019}%
  \BibitemOpen
  \bibfield  {author} {\bibinfo {author} {\bibfnamefont {Sam}\ \bibnamefont
  {McArdle}}, \bibinfo {author} {\bibfnamefont {Xiao}\ \bibnamefont {Yuan}}, \
  and\ \bibinfo {author} {\bibfnamefont {Simon}\ \bibnamefont {Benjamin}},\
  }\bibfield  {title} {\enquote {\bibinfo {title} {{Error-Mitigated Digital
  Quantum Simulation}},}\ }\href {\doibase 10.1103/PhysRevLett.122.180501}
  {\bibfield  {journal} {\bibinfo  {journal} {Phys. Rev. Lett.}\ }\textbf
  {\bibinfo {volume} {122}},\ \bibinfo {pages} {180501} (\bibinfo {year}
  {2019})},\ \Eprint {http://arxiv.org/abs/1807.02467} {arXiv:1807.02467}
  \BibitemShut {NoStop}%
\bibitem [{\citenamefont {Bonet-Monroig}\ \emph {et~al.}(2018)\citenamefont
  {Bonet-Monroig}, \citenamefont {Sagastizabal}, \citenamefont {Singh},\ and\
  \citenamefont {O'Brien}}]{Bonet-Monroig2018}%
  \BibitemOpen
  \bibfield  {author} {\bibinfo {author} {\bibfnamefont {X.}~\bibnamefont
  {Bonet-Monroig}}, \bibinfo {author} {\bibfnamefont {R.}~\bibnamefont
  {Sagastizabal}}, \bibinfo {author} {\bibfnamefont {M.}~\bibnamefont {Singh}},
  \ and\ \bibinfo {author} {\bibfnamefont {T.~E.}\ \bibnamefont {O'Brien}},\
  }\bibfield  {title} {\enquote {\bibinfo {title} {{Low-cost error mitigation
  by symmetry verification}},}\ }\href {\doibase 10.1103/PhysRevA.98.062339}
  {\bibfield  {journal} {\bibinfo  {journal} {Phys. Rev. A}\ }\textbf {\bibinfo
  {volume} {98}},\ \bibinfo {pages} {062339} (\bibinfo {year}
  {2018})}\BibitemShut {NoStop}%
\bibitem [{\citenamefont {Sagastizabal}\ \emph {et~al.}(2019)\citenamefont
  {Sagastizabal}, \citenamefont {Bonet-Monroig}, \citenamefont {Singh},
  \citenamefont {Rol}, \citenamefont {Bultink}, \citenamefont {Fu},
  \citenamefont {Price}, \citenamefont {Ostroukh}, \citenamefont
  {Muthusubramanian}, \citenamefont {Bruno}, \citenamefont {Beekman},
  \citenamefont {Haider}, \citenamefont {O'Brien},\ and\ \citenamefont
  {DiCarlo}}]{Sagastizabal2019}%
  \BibitemOpen
  \bibfield  {author} {\bibinfo {author} {\bibfnamefont {R.}~\bibnamefont
  {Sagastizabal}}, \bibinfo {author} {\bibfnamefont {X.}~\bibnamefont
  {Bonet-Monroig}}, \bibinfo {author} {\bibfnamefont {M.}~\bibnamefont
  {Singh}}, \bibinfo {author} {\bibfnamefont {M.~A.}\ \bibnamefont {Rol}},
  \bibinfo {author} {\bibfnamefont {C.~C.}\ \bibnamefont {Bultink}}, \bibinfo
  {author} {\bibfnamefont {X.}~\bibnamefont {Fu}}, \bibinfo {author}
  {\bibfnamefont {C.~H.}\ \bibnamefont {Price}}, \bibinfo {author}
  {\bibfnamefont {V.~P.}\ \bibnamefont {Ostroukh}}, \bibinfo {author}
  {\bibfnamefont {N.}~\bibnamefont {Muthusubramanian}}, \bibinfo {author}
  {\bibfnamefont {A.}~\bibnamefont {Bruno}}, \bibinfo {author} {\bibfnamefont
  {M.}~\bibnamefont {Beekman}}, \bibinfo {author} {\bibfnamefont
  {N.}~\bibnamefont {Haider}}, \bibinfo {author} {\bibfnamefont {T.~E.}\
  \bibnamefont {O'Brien}}, \ and\ \bibinfo {author} {\bibfnamefont
  {L.}~\bibnamefont {DiCarlo}},\ }\bibfield  {title} {\enquote {\bibinfo
  {title} {{Experimental error mitigation via symmetry verification in a
  variational quantum eigensolver}},}\ }\href {\doibase
  10.1103/PhysRevA.100.010302} {\bibfield  {journal} {\bibinfo  {journal}
  {Phys. Rev. A}\ }\textbf {\bibinfo {volume} {100}},\ \bibinfo {pages}
  {010302} (\bibinfo {year} {2019})},\ \Eprint
  {http://arxiv.org/abs/1902.11258} {arXiv:1902.11258} \BibitemShut {NoStop}%
\bibitem [{\citenamefont {Izmaylov}\ \emph {et~al.}(2019)\citenamefont
  {Izmaylov}, \citenamefont {Yen},\ and\ \citenamefont
  {Ryabinkin}}]{Izmaylov2019}%
  \BibitemOpen
  \bibfield  {author} {\bibinfo {author} {\bibfnamefont {Artur~F.}\
  \bibnamefont {Izmaylov}}, \bibinfo {author} {\bibfnamefont {Tzu-Ching}\
  \bibnamefont {Yen}}, \ and\ \bibinfo {author} {\bibfnamefont {Ilya~G.}\
  \bibnamefont {Ryabinkin}},\ }\bibfield  {title} {\enquote {\bibinfo {title}
  {{Revising the measurement process in the variational quantum eigensolver: is
  it possible to reduce the number of separately measured operators?}}}\ }\href
  {\doibase 10.1039/C8SC05592K} {\bibfield  {journal} {\bibinfo  {journal}
  {Chem. Sci.}\ }\textbf {\bibinfo {volume} {10}},\ \bibinfo {pages}
  {3746--3755} (\bibinfo {year} {2019})},\ \Eprint
  {http://arxiv.org/abs/1810.11602} {arXiv:1810.11602} \BibitemShut {NoStop}%
\bibitem [{\citenamefont {Izmaylov}\ \emph {et~al.}(2020)\citenamefont
  {Izmaylov}, \citenamefont {Yen}, \citenamefont {Lang},\ and\ \citenamefont
  {Verteletskyi}}]{Izmaylov2019b}%
  \BibitemOpen
  \bibfield  {author} {\bibinfo {author} {\bibfnamefont {Artur~F.}\
  \bibnamefont {Izmaylov}}, \bibinfo {author} {\bibfnamefont {Tzu-Ching}\
  \bibnamefont {Yen}}, \bibinfo {author} {\bibfnamefont {Robert~A.}\
  \bibnamefont {Lang}}, \ and\ \bibinfo {author} {\bibfnamefont {Vladyslav}\
  \bibnamefont {Verteletskyi}},\ }\bibfield  {title} {\enquote {\bibinfo
  {title} {{Unitary Partitioning Approach to the Measurement Problem in the
  Variational Quantum Eigensolver Method}},}\ }\href {\doibase
  10.1021/acs.jctc.9b00791} {\bibfield  {journal} {\bibinfo  {journal} {J.
  Chem. Theory Comput.}\ }\textbf {\bibinfo {volume} {16}},\ \bibinfo {pages}
  {190--195} (\bibinfo {year} {2020})},\ \Eprint
  {http://arxiv.org/abs/1907.09040} {arXiv:1907.09040} \BibitemShut {NoStop}%
\bibitem [{\citenamefont {Gokhale}\ \emph {et~al.}(2019)\citenamefont
  {Gokhale}, \citenamefont {Angiuli}, \citenamefont {Ding}, \citenamefont
  {Gui}, \citenamefont {Tomesh}, \citenamefont {Suchara}, \citenamefont
  {Martonosi},\ and\ \citenamefont {Chong}}]{Gokhale2019}%
  \BibitemOpen
  \bibfield  {author} {\bibinfo {author} {\bibfnamefont {Pranav}\ \bibnamefont
  {Gokhale}}, \bibinfo {author} {\bibfnamefont {Olivia}\ \bibnamefont
  {Angiuli}}, \bibinfo {author} {\bibfnamefont {Yongshan}\ \bibnamefont
  {Ding}}, \bibinfo {author} {\bibfnamefont {Kaiwen}\ \bibnamefont {Gui}},
  \bibinfo {author} {\bibfnamefont {Teague}\ \bibnamefont {Tomesh}}, \bibinfo
  {author} {\bibfnamefont {Martin}\ \bibnamefont {Suchara}}, \bibinfo {author}
  {\bibfnamefont {Margaret}\ \bibnamefont {Martonosi}}, \ and\ \bibinfo
  {author} {\bibfnamefont {Frederic~T.}\ \bibnamefont {Chong}},\ }\bibfield
  {title} {\enquote {\bibinfo {title} {{Minimizing State Preparations in
  Variational Quantum Eigensolver by Partitioning into Commuting Families}},}\
  }\href {http://arxiv.org/abs/1907.13623} {\  (\bibinfo {year} {2019})},\
  \Eprint {http://arxiv.org/abs/1907.13623} {arXiv:1907.13623} \BibitemShut
  {NoStop}%
\bibitem [{\citenamefont {Jordan}\ and\ \citenamefont
  {Wigner}(1928)}]{Jordan1928}%
  \BibitemOpen
  \bibfield  {author} {\bibinfo {author} {\bibfnamefont {P.}~\bibnamefont
  {Jordan}}\ and\ \bibinfo {author} {\bibfnamefont {E.}~\bibnamefont
  {Wigner}},\ }\bibfield  {title} {\enquote {\bibinfo {title} {{\"{U}ber das
  Paulische {\"{A}}quivalenzverbot}},}\ }\href {\doibase 10.1007/BF01331938}
  {\bibfield  {journal} {\bibinfo  {journal} {Z. Angew. Phys.}\ }\textbf
  {\bibinfo {volume} {47}},\ \bibinfo {pages} {631--651} (\bibinfo {year}
  {1928})}\BibitemShut {NoStop}%
\bibitem [{\citenamefont {Tranter}\ \emph {et~al.}(2015)\citenamefont
  {Tranter}, \citenamefont {Sofia}, \citenamefont {Seeley}, \citenamefont
  {Kaicher}, \citenamefont {McClean}, \citenamefont {Babbush}, \citenamefont
  {Coveney}, \citenamefont {Mintert}, \citenamefont {Wilhelm},\ and\
  \citenamefont {Love}}]{Tranter2015}%
  \BibitemOpen
  \bibfield  {author} {\bibinfo {author} {\bibfnamefont {Andrew}\ \bibnamefont
  {Tranter}}, \bibinfo {author} {\bibfnamefont {Sarah}\ \bibnamefont {Sofia}},
  \bibinfo {author} {\bibfnamefont {Jake}\ \bibnamefont {Seeley}}, \bibinfo
  {author} {\bibfnamefont {Michael}\ \bibnamefont {Kaicher}}, \bibinfo {author}
  {\bibfnamefont {Jarrod}\ \bibnamefont {McClean}}, \bibinfo {author}
  {\bibfnamefont {Ryan}\ \bibnamefont {Babbush}}, \bibinfo {author}
  {\bibfnamefont {Peter~V.}\ \bibnamefont {Coveney}}, \bibinfo {author}
  {\bibfnamefont {Florian}\ \bibnamefont {Mintert}}, \bibinfo {author}
  {\bibfnamefont {Frank}\ \bibnamefont {Wilhelm}}, \ and\ \bibinfo {author}
  {\bibfnamefont {Peter~J.}\ \bibnamefont {Love}},\ }\bibfield  {title}
  {\enquote {\bibinfo {title} {{The Bravyi-Kitaev transformation: Properties
  and applications}},}\ }\href {\doibase 10.1002/qua.24969} {\bibfield
  {journal} {\bibinfo  {journal} {Int. J. Quantum Chem.}\ }\textbf {\bibinfo
  {volume} {115}},\ \bibinfo {pages} {1431--1441} (\bibinfo {year}
  {2015})}\BibitemShut {NoStop}%
\bibitem [{\citenamefont {Tranter}\ \emph {et~al.}(2018)\citenamefont
  {Tranter}, \citenamefont {Love}, \citenamefont {Mintert},\ and\ \citenamefont
  {Coveney}}]{Tranter2018}%
  \BibitemOpen
  \bibfield  {author} {\bibinfo {author} {\bibfnamefont {Andrew}\ \bibnamefont
  {Tranter}}, \bibinfo {author} {\bibfnamefont {Peter~J.}\ \bibnamefont
  {Love}}, \bibinfo {author} {\bibfnamefont {Florian}\ \bibnamefont {Mintert}},
  \ and\ \bibinfo {author} {\bibfnamefont {Peter~V.}\ \bibnamefont {Coveney}},\
  }\bibfield  {title} {\enquote {\bibinfo {title} {{A Comparison of the
  Bravyi-Kitaev and Jordan-Wigner Transformations for the Quantum Simulation of
  Quantum Chemistry}},}\ }\href {\doibase 10.1021/acs.jctc.8b00450} {\bibfield
  {journal} {\bibinfo  {journal} {J. Chem. Theory Comput.}\ }\textbf {\bibinfo
  {volume} {14}},\ \bibinfo {pages} {5617--5630} (\bibinfo {year} {2018})},\
  \Eprint {http://arxiv.org/abs/1812.02233} {arXiv:1812.02233} \BibitemShut
  {NoStop}%
\bibitem [{\citenamefont {Bravyi}\ and\ \citenamefont
  {Kitaev}(2002)}]{Bravyi2000}%
  \BibitemOpen
  \bibfield  {author} {\bibinfo {author} {\bibfnamefont {Sergey~B.}\
  \bibnamefont {Bravyi}}\ and\ \bibinfo {author} {\bibfnamefont {Alexei~Yu.}\
  \bibnamefont {Kitaev}},\ }\bibfield  {title} {\enquote {\bibinfo {title}
  {{Fermionic Quantum Computation}},}\ }\href {\doibase 10.1006/aphy.2002.6254}
  {\bibfield  {journal} {\bibinfo  {journal} {Ann. Phys.}\ }\textbf {\bibinfo
  {volume} {298}},\ \bibinfo {pages} {210--226} (\bibinfo {year} {2002})},\
  \Eprint {http://arxiv.org/abs/0003137} {arXiv:0003137 [quant-ph]}
  \BibitemShut {NoStop}%
\bibitem [{\citenamefont {McClean}\ \emph
  {et~al.}(2016{\natexlab{b}})\citenamefont {McClean}, \citenamefont
  {Schwartz}, \citenamefont {Carter},\ and\ \citenamefont
  {de~Jong}}]{McClean2017}%
  \BibitemOpen
  \bibfield  {author} {\bibinfo {author} {\bibfnamefont {Jarrod~R.}\
  \bibnamefont {McClean}}, \bibinfo {author} {\bibfnamefont {Mollie~E.}\
  \bibnamefont {Schwartz}}, \bibinfo {author} {\bibfnamefont {Jonathan}\
  \bibnamefont {Carter}}, \ and\ \bibinfo {author} {\bibfnamefont {Wibe~A.}\
  \bibnamefont {de~Jong}},\ }\bibfield  {title} {\enquote {\bibinfo {title}
  {{Hybrid Quantum-Classical Hierarchy for Mitigation of Decoherence and
  Determination of Excited States}},}\ }\href {\doibase
  10.1103/PhysRevA.95.042308} {\bibfield  {journal} {\bibinfo  {journal} {Phys.
  Rev. A}\ }\textbf {\bibinfo {volume} {95}},\ \bibinfo {pages} {042308}
  (\bibinfo {year} {2016}{\natexlab{b}})},\ \Eprint
  {http://arxiv.org/abs/1603.05681} {arXiv:1603.05681} \BibitemShut {NoStop}%
\bibitem [{\citenamefont {Gokhale}\ and\ \citenamefont
  {Chong}(2019)}]{Gokhale2019b}%
  \BibitemOpen
  \bibfield  {author} {\bibinfo {author} {\bibfnamefont {Pranav}\ \bibnamefont
  {Gokhale}}\ and\ \bibinfo {author} {\bibfnamefont {Frederic~T.}\ \bibnamefont
  {Chong}},\ }\bibfield  {title} {\enquote {\bibinfo {title} {{O($N^3$)
  Measurement Cost for Variational Quantum Eigensolver on Molecular
  Hamiltonians}},}\ }\href {http://arxiv.org/abs/1908.11857} {\  (\bibinfo
  {year} {2019})},\ \Eprint {http://arxiv.org/abs/1908.11857}
  {arXiv:1908.11857} \BibitemShut {NoStop}%
\bibitem [{\citenamefont {Verteletskyi}\ \emph {et~al.}(2019)\citenamefont
  {Verteletskyi}, \citenamefont {Yen},\ and\ \citenamefont
  {Izmaylov}}]{Verteletskyi2019}%
  \BibitemOpen
  \bibfield  {author} {\bibinfo {author} {\bibfnamefont {Vladyslav}\
  \bibnamefont {Verteletskyi}}, \bibinfo {author} {\bibfnamefont {Tzu-Ching}\
  \bibnamefont {Yen}}, \ and\ \bibinfo {author} {\bibfnamefont {Artur~F.}\
  \bibnamefont {Izmaylov}},\ }\bibfield  {title} {\enquote {\bibinfo {title}
  {{Measurement Optimization in the Variational Quantum Eigensolver Using a
  Minimum Clique Cover}},}\ }\href {http://arxiv.org/abs/1907.03358} {\ ,\
  \bibinfo {pages} {1--6} (\bibinfo {year} {2019})},\ \Eprint
  {http://arxiv.org/abs/1907.03358} {arXiv:1907.03358} \BibitemShut {NoStop}%
\bibitem [{\citenamefont {Peixoto}(2014)}]{peixoto_graph-tool_2014}%
  \BibitemOpen
  \bibfield  {author} {\bibinfo {author} {\bibfnamefont {Tiago~P.}\
  \bibnamefont {Peixoto}},\ }\bibfield  {title} {\enquote {\bibinfo {title}
  {The graph-tool python library},}\ }\href {\doibase
  10.6084/m9.figshare.1164194} {\bibfield  {journal} {\bibinfo  {journal}
  {figshare}\ } (\bibinfo {year} {2014}),\
  10.6084/m9.figshare.1164194}\BibitemShut {NoStop}%
\bibitem [{\citenamefont {Abraham}\ \emph {et~al.}(2019)\citenamefont
  {Abraham}, \citenamefont {AduOffei}, \citenamefont {Agarwal}, \citenamefont
  {Akhalwaya}, \citenamefont {Aleksandrowicz}, \citenamefont {Alexander},
  \citenamefont {Amy}, \citenamefont {Arbel}, \citenamefont {Arijit02},
  \citenamefont {Asfaw}, \citenamefont {Avkhadiev}, \citenamefont {Azaustre},
  \citenamefont {AzizNgoueya}, \citenamefont {Banerjee}, \citenamefont
  {Bansal}, \citenamefont {Barkoutsos}, \citenamefont {Barron}, \citenamefont
  {Barron}, \citenamefont {Bello},\ and\ \citenamefont {et~al.}}]{Qiskit}%
  \BibitemOpen
  \bibfield  {author} {\bibinfo {author} {\bibfnamefont {H{\'e}ctor}\
  \bibnamefont {Abraham}}, \bibinfo {author} {\bibnamefont {AduOffei}},
  \bibinfo {author} {\bibfnamefont {Rochisha}\ \bibnamefont {Agarwal}},
  \bibinfo {author} {\bibfnamefont {Ismail~Yunus}\ \bibnamefont {Akhalwaya}},
  \bibinfo {author} {\bibfnamefont {Gadi}\ \bibnamefont {Aleksandrowicz}},
  \bibinfo {author} {\bibfnamefont {Thomas}\ \bibnamefont {Alexander}},
  \bibinfo {author} {\bibfnamefont {Matthew}\ \bibnamefont {Amy}}, \bibinfo
  {author} {\bibfnamefont {Eli}\ \bibnamefont {Arbel}}, \bibinfo {author}
  {\bibnamefont {Arijit02}}, \bibinfo {author} {\bibfnamefont {Abraham}\
  \bibnamefont {Asfaw}}, \bibinfo {author} {\bibfnamefont {Artur}\ \bibnamefont
  {Avkhadiev}}, \bibinfo {author} {\bibfnamefont {Carlos}\ \bibnamefont
  {Azaustre}}, \bibinfo {author} {\bibnamefont {AzizNgoueya}}, \bibinfo
  {author} {\bibfnamefont {Abhik}\ \bibnamefont {Banerjee}}, \bibinfo {author}
  {\bibfnamefont {Aman}\ \bibnamefont {Bansal}}, \bibinfo {author}
  {\bibfnamefont {Panagiotis}\ \bibnamefont {Barkoutsos}}, \bibinfo {author}
  {\bibfnamefont {George}\ \bibnamefont {Barron}}, \bibinfo {author}
  {\bibfnamefont {George~S.}\ \bibnamefont {Barron}}, \bibinfo {author}
  {\bibfnamefont {Luciano}\ \bibnamefont {Bello}}, \ and\ \bibinfo {author}
  {\bibfnamefont {Yael Ben-Haim}\ \bibnamefont {et~al.}},\ }\href {\doibase
  10.5281/zenodo.2562110} {\enquote {\bibinfo {title} {Qiskit: An open-source
  framework for quantum computing},}\ } (\bibinfo {year} {2019})\BibitemShut
  {NoStop}%
\bibitem [{\citenamefont {Koch}\ \emph {et~al.}(2007)\citenamefont {Koch},
  \citenamefont {Yu}, \citenamefont {Gambetta}, \citenamefont {Houck},
  \citenamefont {Schuster}, \citenamefont {Majer}, \citenamefont {Blais},
  \citenamefont {Devoret}, \citenamefont {Girvin},\ and\ \citenamefont
  {Schoelkopf}}]{Koch2007}%
  \BibitemOpen
  \bibfield  {author} {\bibinfo {author} {\bibfnamefont {Jens}\ \bibnamefont
  {Koch}}, \bibinfo {author} {\bibfnamefont {Terri~M.}\ \bibnamefont {Yu}},
  \bibinfo {author} {\bibfnamefont {Jay}\ \bibnamefont {Gambetta}}, \bibinfo
  {author} {\bibfnamefont {A.~A.}\ \bibnamefont {Houck}}, \bibinfo {author}
  {\bibfnamefont {D.~I.}\ \bibnamefont {Schuster}}, \bibinfo {author}
  {\bibfnamefont {J.}~\bibnamefont {Majer}}, \bibinfo {author} {\bibfnamefont
  {Alexandre}\ \bibnamefont {Blais}}, \bibinfo {author} {\bibfnamefont {M.~H.}\
  \bibnamefont {Devoret}}, \bibinfo {author} {\bibfnamefont {S.~M.}\
  \bibnamefont {Girvin}}, \ and\ \bibinfo {author} {\bibfnamefont {R.~J.}\
  \bibnamefont {Schoelkopf}},\ }\bibfield  {title} {\enquote {\bibinfo {title}
  {Charge-insensitive qubit design derived from the {Cooper} pair box},}\
  }\href {\doibase 10.1103/PhysRevA.76.042319} {\bibfield  {journal} {\bibinfo
  {journal} {Phys. Rev. A}\ }\textbf {\bibinfo {volume} {76}},\ \bibinfo
  {pages} {042319} (\bibinfo {year} {2007})}\BibitemShut {NoStop}%
\bibitem [{\citenamefont {Chow}\ \emph {et~al.}(2011)\citenamefont {Chow},
  \citenamefont {C{\'{o}}rcoles}, \citenamefont {Gambetta}, \citenamefont
  {Rigetti}, \citenamefont {Johnson}, \citenamefont {Smolin}, \citenamefont
  {Rozen}, \citenamefont {Keefe}, \citenamefont {Rothwell}, \citenamefont
  {Ketchen},\ and\ \citenamefont {Steffen}}]{Chow2011}%
  \BibitemOpen
  \bibfield  {author} {\bibinfo {author} {\bibfnamefont {Jerry~M.}\
  \bibnamefont {Chow}}, \bibinfo {author} {\bibfnamefont {A.~D.}\ \bibnamefont
  {C{\'{o}}rcoles}}, \bibinfo {author} {\bibfnamefont {Jay~M.}\ \bibnamefont
  {Gambetta}}, \bibinfo {author} {\bibfnamefont {Chad}\ \bibnamefont
  {Rigetti}}, \bibinfo {author} {\bibfnamefont {B.~R.}\ \bibnamefont
  {Johnson}}, \bibinfo {author} {\bibfnamefont {John~A.}\ \bibnamefont
  {Smolin}}, \bibinfo {author} {\bibfnamefont {J.~R.}\ \bibnamefont {Rozen}},
  \bibinfo {author} {\bibfnamefont {George~A.}\ \bibnamefont {Keefe}}, \bibinfo
  {author} {\bibfnamefont {Mary~B.}\ \bibnamefont {Rothwell}}, \bibinfo
  {author} {\bibfnamefont {Mark~B.}\ \bibnamefont {Ketchen}}, \ and\ \bibinfo
  {author} {\bibfnamefont {M.}~\bibnamefont {Steffen}},\ }\bibfield  {title}
  {\enquote {\bibinfo {title} {Simple all-microwave entangling gate for
  fixed-frequency superconducting qubits},}\ }\href {\doibase
  10.1103/PhysRevLett.107.080502} {\bibfield  {journal} {\bibinfo  {journal}
  {Phys. Rev. Lett.}\ }\textbf {\bibinfo {volume} {107}},\ \bibinfo {pages}
  {080502} (\bibinfo {year} {2011})}\BibitemShut {NoStop}%
\bibitem [{\citenamefont {Wood}\ \emph {et~al.}(2015)\citenamefont {Wood},
  \citenamefont {Biamonte},\ and\ \citenamefont {Cory}}]{Wood2015}%
  \BibitemOpen
  \bibfield  {author} {\bibinfo {author} {\bibfnamefont {Christopher~J.}\
  \bibnamefont {Wood}}, \bibinfo {author} {\bibfnamefont {Jacob~D.}\
  \bibnamefont {Biamonte}}, \ and\ \bibinfo {author} {\bibfnamefont {David~G.}\
  \bibnamefont {Cory}},\ }\bibfield  {title} {\enquote {\bibinfo {title}
  {{Tensor networks and graphical calculus for open quantum systems}},}\
  }\href@noop {} {\bibfield  {journal} {\bibinfo  {journal} {Quantum
  Information and Computation}\ }\textbf {\bibinfo {volume} {15}},\ \bibinfo
  {pages} {759--811} (\bibinfo {year} {2015})},\ \Eprint
  {http://arxiv.org/abs/1111.6950} {arXiv:1111.6950} \BibitemShut {NoStop}%
\bibitem [{\citenamefont {Cjwood}(2019)}]{Cjwood2019}%
  \BibitemOpen
  \bibfield  {author} {\bibinfo {author} {\bibnamefont {Cjwood}},\ }\href@noop
  {} {\enquote {\bibinfo {title} {{How good is
  basic{\_}device{\_}noise{\_}model() simulating the noise in the quantum
  computer?}}}\ } (\bibinfo {year} {2019}),\ \bibinfo {note}
  {https://quantumcomputing.stackexchange.com/questions/8958/how-good-is-basic-device-noise-model-simulating-the-noise-in-the-quantum-compu,
  accessed November 4th, 2020}\BibitemShut {NoStop}%
\bibitem [{\citenamefont {Nielsen}\ and\ \citenamefont
  {Chuang}(2010)}]{Nielsen2010}%
  \BibitemOpen
  \bibfield  {author} {\bibinfo {author} {\bibfnamefont {Michael~A.}\
  \bibnamefont {Nielsen}}\ and\ \bibinfo {author} {\bibfnamefont {Isaac~L.}\
  \bibnamefont {Chuang}},\ }\href {\doibase 10.1017/CBO9780511976667} {\emph
  {\bibinfo {title} {Cambridge University Press}}}\ (\bibinfo  {publisher}
  {Cambridge University Press},\ \bibinfo {address} {Cambridge},\ \bibinfo
  {year} {2010})\BibitemShut {NoStop}%
\end{thebibliography}%

\end{document}